\documentstyle[aas2pp4,tighten,epsfig]{article}

\newcommand\eg{e.g. }
\newcommand\degr{$^\circ$ }

\newcommand{\Td}{T$_{\rm d}$}
\newcommand{\ha}{H$\alpha$}
\newcommand{\hi}{H{\sc i}}
\newcommand{\hii}{H{\sc ii}}
\newcommand{\Sigh}{$\Sigma_{\rm H}$}
\newcommand{\Sighone}{$\Sigma_{\rm{HI}}$}
\newcommand{\Sightwo}{$\Sigma_{{\rm H}_2}$}
\newcommand{\Sigdust}{$\Sigma_{\rm d}$}
\newcommand\tausixty{$\tau_{60}$}
\newcommand{\nh}{N$_{\rm{H}}$}

\newcommand{\htwo}{H$_2$}
\newcommand{\tauvsixty}{$\tau_{\rm V}/\tau_{60}$}

\begin{document}
\baselineskip=15pt      

\title{Spatial distribution of Far infrared emission in spiral galaxies
II. Heating sources and gas-to-dust ratio}
\author{Y.D.\,Mayya{$^{1,2}$} and T.N.\,Rengarajan$^1$}
\affil{$^1$Tata Institute of Fundamental Research, Homi Bhabha Road, 
        Mumbai 400 005, India}  
\affil{$^2$Instituto Nacional de Astrofisica Optica y Electronica,  
           Apdo Postal 51 y 216, 72000 Puebla, Pue., M\'EXICO   }   
\affil{Electronic Mail: ydm@inaoep.mx  and renga@tifrvax.tifr.res.in}
\authoremail{ydm@inaoep.mx  and renga@tifrvax.tifr.res.in}
\vskip 0.5cm
\affil{\it Accepted --- June 1997. 
To appear in Astronomical Journal, September 1997}


\begin{abstract}
We study the radial distribution of the temperature of the warm dust and 
gas-to-dust mass ratios in a sample of 22 spiral galaxies. 
The heating capabilities of the diffuse interstellar radiation field (ISRF), 
based on D\'esert et al. model, are investigated in 13 of the sample galaxies.
In general, the temperature of the warm dust decreases away from the center,
reaches a minimum value at the mid-disk and increases again 
in the outer parts of galaxies. Heating a mixture of small and big grains 
by the ISRF is able to explain the observed behavior qualitatively.
However, ultraviolet photons from recent star formation events are 
necessary for a detailed matching of the warm dust temperature profiles.
Very small grains contribute typically more than 50\% to the observed flux 
at 60\micron\ beyond half the disk radius in galaxies.
Optical depth (\tausixty) profiles, derived from the observed 60\micron\ and
warm dust temperature profiles, peak at or close to the galactic center. 
In 13 of the galaxies, where dust temperature profiles are modeled, we obtain 
optical depth and dust mass profiles after correction for the contaminating 
effects of very small grains. These profiles are combined with the gas density 
profiles in the literature, to generate profiles of the gas-to-dust mass ratio.
The resulting gas-to-dust mass ratio decreases by a factor of 8 from the center 
to the optical isophotal radius, where the value approaches the local galactic 
value. With the understanding that the dust mass is proportional to 
metallicity, and that the metallicity increases towards the center of galaxies, 
one expects the 
gas-to-dust ratio to decrease towards the center, contrary to what 
is observed. We demonstrate that the observed steep gradient is a result of 
the over-estimation of the molecular mass, and can be flattened out to within a 
factor of 2, if the molecular hydrogen mass is recomputed assuming a metallicity 
dependent conversion factor from CO intensity to \htwo\ column density. 
The flattened radial profiles indicate a global gas-to-dust ratio of around 
300, which is within a factor of two of the local galactic value.

\keywords{dust -- extinction -- infrared radiation -- heating sources -- 
                  gas-to-dust ratio}

\end{abstract}

\section{Introduction}    

Infrared Astronomical Satellite (IRAS)
observations have made a significant impact on our understanding of the 
dust content in galaxies. These observations were mainly sensitive to dust 
with temperature in the range 25--50~K, generally referred to as warm 
dust. There have been considerable debates on the heating sources of dust 
(Helou et al. 1985; Rengarajan \& Verma 1986; Devereux et al. 1994; 
Walterbos \& Schwering 1987; Xu \& Helou 1996; Walterbos \& Greenawalt 1996)
 and the mass contained in 
the cooler dust (Devereux \& Young 1990; Chini \& Kr\"ugel 1993).
The ultraviolet (UV) photons from young hot stars are efficient in heating the 
dust and hence the most luminous far infrared (FIR) galaxies are inferred to 
be in a starburst mode. Such galaxies are characterized by high ratio of 
FIR to blue band luminosity. 
The heating provided by the non-ionizing photons from young and old
stars may play an important role in regions of low levels of
current star formation activity, as characterized by the low values of FIR 
to blue band luminosity. Investigating the heating power of the
non-ionizing photons in these low-luminosity galaxies is one of the main
goals of this study. From studies carried out in M\,31, 
Devereux et al. (1994) find that about 30\% of the FIR luminosity comes from 
regions away from current star formation, whereas Walterbos \& Schwering 
(1987) and more recently Xu \& Helou (1996) and 
Walterbos \& Greenawalt (1996) infer even higher levels of 
heating provided by the old disk and bulge stars. 

Dust and gas in the interstellar medium (ISM) are believed to be well mixed, 
with a gas-to-dust mass ratio of around 160 in the local galactic 
surroundings (Sodroski et al. 1994). Based on global data of galaxies, 
Devereux \& Young (1990) derive a value as high 
as 1000 for the gas-to-dust mass ratio in external galaxies. 
Considering that the IRAS observations are insensitive to the emission 
from dust below 25~K, the inferred high gas-to-dust ratio
could be the result of vast amounts of dust mass in cooler dust.
However observations carried out at longer wavelengths to trace the 
cool dust have given inconclusive results regarding the amount of dust 
in the cool component (see Chini \& Kr\"ugel 1993). Andreani et al. 
(1995) used millimeter (mm) continuum observations 
to infer emission from cool dust,
and found gas-to-dust mass ratio to lie in the range 100--1000 with a 
mean value around 230, which is close to the local galactic value.

In a companion paper (Mayya \& Rengarajan 1997; Paper I), we have studied
the spatial distribution of FIR and 20~cm radio continuum (RC) emission
in a sample of 22 galaxies. The galaxies were chosen with a view to study
the spatial variation of warm dust temperature as well as the gas-to-dust ratio 
within the galaxies. In this paper, we combine the FIR data with the 
published gas density profiles for all the galaxies and optical data for a 
subsample of 13 galaxies, and address the issues of dust heating and 
gas-to-dust ratios in galaxies. Sec.~2 deals with the data and the
analysis procedure that we have followed. Observed gradients in 
the warm dust temperature are compared with the dust heating models in
Sec.~3. In Sec.~4, we discuss the gradients in FIR optical depth and the
gas-to-dust ratios in galaxies. Concluding remarks are made in Sec.~5.

\section{ The sample, data and analysis procedure}   

Galaxies for study here, as well as in Paper~I, were selected 
based on the following criteria.
(1) The availability of 20~cm radio continuum images at 1\arcmin\ resolution,
(2) the availability of \hi\ and \htwo\ radial surface density profiles,  
(3) the optical angular diameter $>5$\arcmin\ and
(4) the 60\micron\ flux density  $>5$~Jy.
The last two criteria together ensure that there are
significant number of resolution elements in each image, and that the
galaxies are bright enough in the IRAS bands to get reliable High Resolution
(HiRes) images. 
Among the 22 galaxies studied in Paper~I, NGC\,4656 does not have published 
radial profiles of \hi\ and \htwo\ surface densities, and is replaced by 
NGC\,4258 in the present work.

The basic properties of the sample galaxies are given in Table~1. The galaxy
type, the optical diameter in arcmin at 25~$B$~mag\,arcsec$^{-2}$, the axis 
ratio {\it b/a} and the position angle (PA) of the major axis
are taken from the {\it Third reference catalogue of bright galaxies} 
(de Vaucouleurs et al. 1991; RC3 henceforth). Distances
are taken from the Nearby Galaxy Catalog (Tully 1988), where 
a Hubble constant of 75~km\,s$^{-1}$\,Mpc$^{-1}$ is assumed. 
The total (1--1000\micron) FIR luminosity (L$_{\rm fir}$) to optical blue 
band luminosity (L$_{\rm B}$) ratio is calculated using the formula,
\begin{equation}
{{\rm L}_{\rm fir}\over {\rm L_{\rm B}}} = 
{1.26\times 10^{-11} C (2.58 \rm S_{60} + \rm S_{100})\over
{5.98\times 10^{-6} 10^{-0.4 \rm B_{\rm T}^0}}},
\end{equation}
where $C$ corrects for the flux outside the 60 and 100\micron\ bands and is
tabulated by Helou et al. (1988) as a function of temperature for a 
single temperature model. For the observed range of dust temperatures
in galaxies, $C$ lies between 1.3--2.0 with a typical value of 1.5. 
L$_{\rm B}$ is the true in-band $B$ luminosity.
We note that Young et al. (1989) transformed magnitude 
B$_{\rm T}^0$ to fluxes using the absolute $B$ magnitude of the Sun 
and hence their formula results in the over-estimation of L$_{\rm B}$ 
and hence the under-estimation of the ratio L$_{\rm fir}$/L$_{\rm B}$
by a factor L$_\odot$/L$_{{\rm B}_\odot} = 9$.
S$_{60}$ and S$_{100}$ are the global 60\micron\ and 100\micron\ flux 
densities (in units of Jy) respectively. The denominator and
numerator are in units of erg\,cm$^{-2}$\,s$^{-1}$. 
The sources of \hi\ and \htwo\ data appear in the last two columns of
Table~1 respectively. Note that the \htwo\ data is unavailable for NGC\,3198. 
For galaxies with known metallicity gradients 
(Zaritsky et al. 1994), the oxygen abundance at 3~kpc and its gradient in 
dex/kpc are given. Thirteen of the galaxies have archival CCD data available 
at broad optical bands, based on observations at the Palomar Observatory 
(Frei et al. 1996). These CCD images 
are ideal in the investigation of dust heating by the non-ionizing photons.

We use the HiRes images (Rice 1993) of the galaxies at the four IRAS bands
in this work. The HiRes resolutions at 60\micron\ in arcsec along two 
perpendicular directions  are given in Table~1. The
resolution is better at 12 and 25\micron\ bands, but the sensitivities in 
these bands are too poor to trace out the outer regions in most of the
galaxies in our sample and hence these images have only limited use in this 
study. The HiRes processing was done
under the default configuration, which results in frames of 
$1^\circ\times1^\circ$\ field with 15\arcsec\ pixels. Sub-images of
60$\times$60 pixels are extracted for all galaxies except NGC\,2403 and
4258, for which sections of size 100$\times$100 pixels are extracted. 
In all cases, images from the 20th iteration are used. Contour diagrams of
program galaxies at 60\micron\ are presented in Paper~I.

Surface brightness profiles of HiRes images are obtained by 
azimuthally averaging over elliptical annuli of 1\arcmin\ width. 
The ellipticity and position angle of the
annuli are fixed at their optical value as tabulated in Table~1.
These radial profiles are used to obtain the S$_{60}$ to S$_{100}$ color
profile, which is transformed into a dust temperature profile using the
$\lambda^{-1}$ emissivity law and the Planck's function. We refer to the 
resultant profiles as warm dust profiles. The physical temperature of the dust 
in any region in a galaxy depends on the
type and size distribution of dust particles, 
effective wavelength of the heating source and the distance of the 
heating source from the dust particles. 
As all these quantities can
vary within a single annulus, the derived warm dust temperature is
expected to match the local dust temperature only if there is a single
type of grain. In reality, the ISM contains a mixture
of different grain types, and hence, the derived warm
dust temperature is a flux-weighted mean of all the grain temperatures.

The optical depth of the FIR emitting dust at 60\micron\ ($\tau_{60}$) is 
defined by the equation, 
\begin{equation}
 \tau_{60}  = {S_{60}\over{B(60,{\rm T}_{\rm d})}}, 
\end{equation}
for the optically thin case ($\tau_{60} << 1$). $S_{60}$ and $B(60,T_d)$ 
in the above equation are respectively the observed intensity within the 
60\micron\ band and the Planck function at the estimated temperature \Td\ 
and wavelength 60\micron. Observed warm dust temperature profiles and the 
60\micron\ surface brightness profiles are used in obtaining $\tau_{60}$
profiles. 
$\tau_{60}$ is transformed into a dust mass surface 
density \Sigdust\ using the formula 
\begin{equation}
 \Sigma_{\rm d} ({\rm M}_\odot {\rm pc}^{-2}) = 108\tau_{60},
\end{equation}
where the constant 108 is a measure of the dust size distribution and the
scattering properties of the dust, and is computed based on the constants
in Hilderbrand (1983) corresponding to the dust properties in the solar
neighborhood. Classical big dust grains control the dust mass, where as 
small grains when present in an annulus, control \Td.  
Thus for the purposes of deriving dust mass, it is necessary to use the 
values of \Td\ appropriate to big grains only in Eq.~2. Details of the
procedure followed in estimating such a \Td\ are discussed in Sec.~3.3.\\
The routines under Image Reduction and Analysis Facility (IRAF) and Space 
Telescope Science Data Analysis System (STSDAS) are used in the reduction and 
the analyses of the data. 

\section{Distribution of warm dust temperature }   

\subsection{Radial profiles}      

As described in the previous section, the warm dust temperature (\Td) profiles 
are obtained from 60 and 100\micron\ profiles assuming a $\lambda^{-1}$
emissivity law. 
Resultant radial profiles for all the 22 galaxies are given in Fig.~1. 
Radial distances are expressed in units of optical isophote radius
(R$_{25}$). The same temperature range is used in all the
plots to enable easy comparison of absolute values as well as gradients
from galaxy to galaxy. The solid line passing through the circles denotes 
the azimuthally averaged profiles, sampled at 1\arcmin\ intervals.
The horizontal dotted line denotes the mean dust 
temperature of the galaxies as derived from the global data.
The following general trends can be noticed from these profiles.\\
1. In the inner half of the galaxy, the warm dust temperature decreases 
   as one goes radially outwards from the center in a majority of the 
   galaxies. The decrease is less than 3~K in galaxies 
   NGC\,628, 2403, 2841 and 7331, and more than 10~K in galaxies 
   NGC\,2903, 3079, 3628, 4258, 4535, 4736, 5033 and 6946. In the 
   remaining 10 galaxies, \Td\ decreases by around 5--10~K. \\
2. Beyond half the disk radius, \Td\ shows a tendency to increase 
   in about 50\% of the cases. \\
3. The warm dust temperatures derived from global data are nearly the 
    same as those for the high intensity inner regions. \\ 
  
The slightly poorer resolution of the 100-\micron\ images as compared to
the 60-\micron\ images, may result in over-estimation of the central \Td. 
However the result (3) above, stating that the central \Td\ is not very much 
higher than the global \Td, indicates that the \Td\ is not severely
affected by the difference in resolution of the two images. This 
indicates that the bulk of the FIR emission originates in regions
which are much smaller than the beams in both the bands. This is consistent
with the picture that the nuclear star forming regions and/or
the bulge stars power the FIR emission.

\subsection{Discussion on the heating sources}  

There has been a lot of discussion regarding the nature of heating sources
of dust traceable by IRAS observations (see \eg\ Devereux et al. 1994). 
The debate is on the fraction of
heating provided by the non-ionizing photons from low to intermediate mass
stars, which populate the disk of galaxies as against the heating by the 
Lyman continuum photons from the massive stars. Recent studies of M\,31,
M\,101 and M\,81 have revealed good positional coincidences between the \ha\ 
and FIR emitting sources (Devereux et al. 1994; Devereux \& Scowen 1994;
Devereux et al. 1995), indicating that the massive stars when present act as 
efficient heating sources. The 1\arcmin\ resolution in the
present study corresponds to a linear size of 1--6~kpc in
the program galaxies, which is larger than the extent of a typical star 
forming complex. Thus the observed \Td\ corresponds to a mixture of 
several populations. However the vastly differing
heating efficiencies of ionizing and non-ionizing photons can be used to
study the heating capabilities of old stars, 
which is the topic of discussion in the remaining part of this section. 

We first compute the amount of heating  by the 
interstellar radiation field (ISRF), 
which is mainly originating from long-lived low and intermediate mass stars. 
We use the calculations of D\'esert et al. (1990), who computed the 
expected flux densities within the IRAS bands for an assumed dust
composition and spectrum of the heating radiation field. The dust 
composition chosen was such that it reproduces the galactic interstellar 
extinction curve (Savage \& Mathis 1979), and contains
Very Small Grains (VSGs), Polycyclic Aromatic Hydrocarbons (PAHs) in
addition to the conventional Big Grains (BGs). They assumed the spectrum of 
the heating radiation field to be similar to that of the local
interstellar radiation field (LISRF). The resulting flux densities in the
IRAS bands (normalized to total hydrogen column densities) were tabulated as a
function of the radiation field intensity in the $V$-band for each
component of dust. They also computed the IRAS flux densities for
ultraviolet enriched (with respect to LISRF) input radiation field
by adding to the LISRF a component denoted as $X_{\rm uv}$ ($X_{\rm uv}=0$ 
and 3 correspond to LISRF and 3 times UV enriched LISRF respectively). 

The models of D\'esert et al.  can be used to compute the expected \Td\ 
gradients in galaxies, if $V$-band surface brightness profiles are known. 
For 13 of
the program galaxies such data are available in a recently compiled data
base (Frei et al. 1996), based on the observations at the Palomar 
Observatory. We assumed their CCD $g$-band ($\lambda 5000/700$) images to
represent the $V$-band surface brightness, and obtained $V$-band surface 
brightness profiles using methods identical to that followed in extracting
60 and 100\micron\ profiles (Sec.~2). The ISRF as seen by the dust particles 
differs from the observed $V$-band surface brightness, due to the extinction 
by dust within the parent galaxy. The amount of visual extinction can be 
estimated using the observed gas density (\nh) along with the galactic value
for A$_{\rm v}$/\nh\ (A$_{\rm v}/{\rm N}_{\rm H}=5.3\times 10^{-22}$ cm$^2$: 
Bohlin et al. 1978). Using A$_{\rm v}$ to estimate the ISRF from the observed 
optical surface brightness may be an upper limit as some of the
scattered photons recontribute to the observed surface brightness. Also
only those dust particles in the foreground of stars will contribute to
extinction and hence effectively only half the amount of particles 
might be contributing to extinction. Accordingly, we correct the $V$-band
surface brightness profiles by A$_{\rm v}$ as estimated from \nh/2. 
\nh\ includes both the atomic and molecular contributions and were taken 
from the sources listed in Table~1.

The modeled \Td\ profiles for the 13 galaxies with optical CCD data are 
plotted in 13 panels of Fig.~2, along with the azimuthally averaged 
observed profiles. Open circles represent the observed \Td\ profile. 
Four models are shown by different line types which are identified in 
the last panel. Three of these correspond to the same dust mixture 
(VSG$+$PAH$+$BG), but for three different heating spectra, while the 
fourth one (dashed line) corresponds to the case where the ISM contains 
only the classical big grains. The solid and dash-dotted lines use the 
LISRF spectrum with ($X_{\rm uv}=3$) and without ($X_{\rm uv}=0$) UV 
enrichment respectively. For these two models the observed $V$-band 
profiles are corrected for the extinction as described above. For the 
third model (dotted line) $X_{\rm uv}=3$, but the observed $V$-band 
surface brightness profile is used (A$_{\rm v}=0$). The heating spectrum 
for the fourth model is identical with the third model and hence the 
contribution of the VSGs$+$PAHs (which increases with radius) can be 
inferred by comparing the dotted and dashed lines. On the other hand, 
the difference between the solid and dotted lines illustrate the effect 
of the extinction, which is found to be significant in the central 
regions of galaxies. The difference between the solid line and the 
dash-dotted line (which is around 3--4~K) shows the effect of enhancement 
in the UV part of the heating spectrum. 

With the chosen range of parameters related to the dust composition, heating 
radiation field and extinction, models encompass the observed range of \Td\ in
a majority of the galaxies. We now select the models which best match the 
observed \Td\ profile in individual galaxies. The most obvious trend in the
observed \Td\ profiles is the initial fall and the outward increase, which 
is fairly well reproduced by the models having a dust mixture containing 
PAHs and VSGs in addition to BGs. With only BGs, \Td\ is expected to 
monotonically decrease with distance from the center in all the galaxies. 
VSGs control the \Td\ in the outer disk, where
the radiation intensity is low, whereas BGs do the same in the inner disk.
However the amount of heating provided by the old stars with a spectrum 
similar to that of the LISRF ($X_{\rm uv}=0$) is found to be adequate to 
heat the dust to the observed \Td\ only in NGC\,4501 and 5055.
In the remaining galaxies, better results are obtained by using a spectrum with an
enhanced UV continuum, which mimics the addition of young massive stars. 
It is expected that the UV contribution is not the same at all radial 
distances. However, we use a single $X_{\rm uv}$ value for all radii
in order to show the role of this parameter in heating the dust.
UV enhancement of a factor of three is found to be sufficient at most
radial positions in seven of the remaining eleven galaxies (NGC\,2903, 
4254, 4258, 4303, 4321, 4535, 5033) studied here. In NGC\,2403 \& 3198, the 
observed \Td\ is around 5~K hotter than the ISRF
heating models with $X_{\rm uv}=3$ in most parts of the disks. 
The observed \Td\ value in the central region of NGC\,4535 is around 8~K
hotter that the $X_{\rm uv}=3$ model, implying the necessity of an
additional heating source. The observed \Td\ is high in the outer 
regions of NGC\,4569. The errors on \Td\ are also expected to be higher in
this galaxy due to the combined effects of lower flux and poorer resolution
compared to other galaxies in the sample. Thus the higher \Td\ in
NGC\,4569 may not be physically significant.

$X_{\rm uv}=3$ corresponds to a value less than 0.5\% of stellar mass in 
the young stars as compared to the old stars per surface area, for a galaxy 
like the Milky Way as modeled by Bruzual \& Charlot (1993). For a typical 
disk surface density of 10$^7$M$_\odot$\,kpc$^{-2}$, this corresponds to 
$5\times10^4$ M$_\odot$\,kpc$^{-2}$ in young stars. Giant \hii\ regions 
in galaxies have typically a mass of $10^5$~M$_\odot$ in
less than 1 kpc$^2$ area (Mayya \& Prabhu 1996). Thus $X_{\rm uv}=3$ can
be easily satisfied locally where giant \hii\ regions are present. However, 
the observed number of giant \hii\ regions in galaxies is insufficient 
to maintain a global $X_{\rm uv}=3$. The heating provided by the ionizing 
photons and Lyman alpha photons, which are not accounted by the
the $X_{\rm uv}$ parameter in D\'esert et al., are essential to explain
the global heating.

Models with the extinction corrected ISRF predict a steep increase of
\Td\ towards the center, which matches well the observed \Td\ 
gradients in NGC\,2903 and 4192. In most of the remaining galaxies, the 
estimated extinction over-corrects the ISRF, thus resulting 
in higher model values of \Td\ in the central regions. While models
without the UV contributing massive stars in the central regions can 
explain the central \Td\ gradients in some cases, the possible range of
uncertainties in the estimation of A$_{\rm v}$ can as well reproduce the
observed flatter gradients. We here point out that the extinction
estimation in the central regions based on the observed gas density, which
is dominated by \htwo, is likely to be uncertain due to our poor
understanding of the dependence of gas-to-dust ratio, and the CO intensity to
molecular hydrogen mass conversion factor, with metallicity. Fortunately,
the two dependences are in the opposite directions and hence 
only the difference in their degree of dependences affects the extinction
estimation. These issues are re-addressed later in Sec.~4.2.

It is interesting to note that the dust heating models with a spectrum 
similar to that of LISRF correctly reproduce the observed shape of the \Td\ 
gradient, though it is cooler by around 5 K in majority of the galaxies. 
We found that, with the dust mixture assumed by D\'esert et al., 
the required additional heating can be provided by the 
UV photons from young stars in giant \hii\ regions. However there are a
few cases, like the outer region in NGC\,2903, and over most of
the radial extent in NGC\,2403 and 3198, where the assumed range of
parameters do not produce the observed \Td\ gradients. Increasing
$X_{\rm uv}$ beyond 3 may result in a better match, but it is not
consistent with the lack of massive stars in the outer region of
NGC\,2903. On the other hand it is likely that the VSG content
in the dust mixture varies from region to region in galaxies and also 
from galaxy to galaxy.  For e.g. Xu \& Helou (1994) 
have found deficiency of VSGs in the ISM of M\,31, compared 
to that used by D\'esert et al. The observed higher \Td\ in the
outer regions of the above galaxies may be due to a higher VSG content. 
Accordingly, we doubled the flux contributed by the VSGs at 60 \& 
100 \micron, and recomputed the resulting \Td\ for the new dust mixture 
in galaxies NGC\,2403, 2903 and 3198. The results of this study 
are plotted in Fig.~3. The open circles and the solid line
have the same meaning as in Fig.~2. The dotted line has the same
parameters as for the solid line, but for the enhanced VSG content.
The plots illustrate that the high observed values of \Td\ can be
explained in terms of enhanced VSG content in the dust mixture. 
Thus we conclude that the enhancement of the VSG content over that used by
D\'esert et al. is required at least in some galaxies in the sample.
More elaborate models, incorporating the heating provided by the ionizing 
photons as estimated from the observed \ha, in addition to the heating by the 
ISRF are required to reproduce the observed \Td\ distribution for individual 
galaxies at all radii. Such studies will be instructive on the
possible variations in the relative amounts of the three 
components of dust in the ISM. However, we continue to use the dust 
composition used by D\'esert et al. in the remaining part of this study.

\subsection{Contribution from big grains alone to the computed quantities} 

For some applications (such as the optical depth, to be discussed in the 
next section) it is important to get an estimate of the contributions 
from big grains alone in the IRAS bands.
Our analysis of the previous sub-section, based on the D\'esert et al. (1990)
models, allows us to compute this contribution. 
We obtain the radial gradients of the 60 and 100\micron\ 
emission expected from the big grains alone in five of the largest 
(in angular size) galaxies in our sample. The results are plotted in the
top two panels of Fig.~4, as fractions of the total (BG$+$VSG$+$PAH) 
emission in the
respective bands. In obtaining these fractions, we have assumed
the $X_{\rm uv}=3$ models after correcting for extinction based on
estimated A$_{\rm v}$ (solid line in Fig.~2). It is interesting to note 
that the radial dependence is similar in the five galaxies studied, 
and hence the results from these plots may be extended to other galaxies too.
The big grain contribution to the 60\micron\ emission falls below 50\%
beyond half the disk radius, reaching values as low as 20\% at 
0.8R$_{25}$. In comparison the 100\micron\ emission
is less affected by emission from VSGs$+$PAHs, contributing 
in the range of 15--25\% beyond half the disk radius.
As we saw in the previous sub-section, this differential contribution
to the 60 and 100\micron\ bands affects the derived dust temperature
the most. In the third panel of Fig.~4, we plot the \Td\ expected
from the big grains alone as a fraction of the observed \Td\ for the same
five galaxies as in the earlier panels. Again each of the five galaxies
show very similar behavior. In the last panel a similar plot is
given for \tausixty(BG)/\tausixty(Obs), where the numerator 
is computed using \Td(BG) and S$_{60}$(BG) in Eq.~2.
In the outer regions, the observationally derived \tausixty\ is an order 
of magnitude lower compared to that expected from the big grains alone. 
This is due to the VSGs$+$PAHs affecting the \tausixty\ indirectly through 
the steep \Td\ dependence (see Eq.~2). \tausixty(BG) can be obtained from 
the observationally derived \tausixty\ by applying mean correction
factors of $-0.05$, 0.08, 0.37, 0.80, 1.25 dex at radii 0.10, 0.25, 0.50, 
0.75 and 1.00 R$_{25}$ respectively. 
These values will be used to estimate \Sigdust\ 
without the contaminating effects of VSGs and PAHs.

\section{Distribution of $\tau_{60}$ and implications}  

At the FIR wavelengths the dust is optically thin to its own radiation and 
hence
the amount of dust along the line of sight can be estimated from the observed
FIR flux densities. Eqs.~2 and 3 in Sec.~2 define the 60\micron\ optical 
depth, $\tau_{60}$, and the dust surface mass density, $\Sigma_{\rm d}$. 
The radial distribution of dust mass and the gas-to-dust ratio in galaxies 
are the subject of discussion in this section.

\subsection{ Radial gradients}  

$\tau_{60}$ profiles for all the galaxies are computed following Eq.~2,  
making use of the azimuthally averaged 60\micron\ intensity profiles
and \Td\ profiles discussed in the previous section.
The resultant profiles (in log units) are plotted in Fig.~5 as a function 
of the radial distance from the center in units of optical disk radius. 
The open circles denote the observed \tausixty\ profile, whereas the
filled circles (in 13 galaxies where we have modeled the \Td\ gradient) 
represent the expected \tausixty\ from big grains alone, following the 
discussions in Sec.~3.3. It should be noted that
VSG and PAH contribute very little to the derived dust mass density; it is
mostly \tausixty(BG) that is a measure of the dust mass density,
and thus we refer \tausixty(BG) as the corrected \tausixty\ henceforth.
The various curves plotted in Fig.~5
correspond to the values estimated from the observed gas density 
profile, which are discussed in detail below.

The observed \tausixty\ in general increases as radius decreases;
in the central region it shows a tendency to flatten off or dip downwards. 
Such trends were also noticed earlier by Ghosh et al. (1993) in a study 
based on maximum entropy deconvolution of IRAS images.
The observed \tausixty\ profiles become flatter after correcting them
using D\'esert et al. model. There are two galaxies --- NGC\,4192 and 
4501 in which the corrected \tausixty\ increases with radius. 
Subtraction of the VSG and PAH contribution based on the model of 
D\'esert et al. (1990) are over-corrections for these galaxies as can be 
inferred from the \Td\ plots (Fig.~2).  
The actual values may be intermediate to the two sets.
It may, however, be noted that the observed \tausixty\ profile 
increases with radius in NGC\,3628 and is almost flat in NGC\,3079. 
These are the only two sample galaxies which are within 15\degr\ of being
edge-on, and may need a separate treatment. 

In order to understand the general behavior of \tausixty\ decreasing 
radially outwards, we look into the various factors controlling the 
optical depth at FIR wavelengths. For a given value of the gas-to-dust 
ratio, \tausixty\ mainly depends on the scattering
properties of the dust mixture and total column density of the gas (\nh). 
It is likely that the gas-to-dust ratio itself depends on the metallicity ($Z$) 
in the galaxy. Thus the optical depth at a FIR wavelength $\lambda$ 
can be written as, 
\begin{equation}
\tau_\lambda = 0.921\left({{\rm A}_{\rm v}\over{\rm N}_{\rm H}}\right)
\left({Z\over{Z_\odot}}\right)^b {\rm N}_{\rm H} 
\left({{\rm A}_\lambda\over{{\rm A}_{\rm v}}}\right),
\end{equation}
where $b$ is the power law index to take into account of a possible dependence 
of the gas-to-dust ratio with metallicity.
Substituting the galactic value of A$_{\rm v}$/\nh, and expressing \nh\ in 
terms of the mass surface density \Sigh, we get,
\begin{equation}
\tau_{60}= 3.66\times10^{-2}\left({{\Sigma_{\rm H}}\over{M_\odot 
{\rm{pc}}^{-2}}}\right).\left({Z\over{Z_\odot}}\right)^b 
\left({\tau_{60}\over{\tau_{\rm V}}}\right),
\end{equation}
where ${\tau_{60}\over{\tau_{\rm V}}}$ is the relative optical depth, which 
can vary over a large range depending on the shape of the extinction curve.
Casey~(1991) has determined the \tauvsixty\ values to be in the range
800--4000 for clouds in the vicinity of hot stars and 150--700 towards
cooler molecular clouds. The effective \tauvsixty\ is determined by the 
size distribution of the dust particles and the effective wavelength of the
radiation field. 

The computation of $\tau_{60}$ requires a
knowledge of the gradients in the gas surface density, metallicity and
\tauvsixty\ as well. The gas surface densities are dominated by the
molecular hydrogen, as traced by the CO line in the central regions and the
neutral hydrogen in the outer regions. Observational data both in atomic
and molecular phases are available for the program galaxies in the literature.
The last two columns in Table~1 list the sources of these data.
The CO line intensity to \htwo\ mass conversion factor is expected to 
be a function of Z; at present we follow the conventional method,
but consider the Z-dependence later in Sec.~4.2. 
Metallicity decreases radially outwards in galaxies.
In a recent study, Zaritsky et al. (1994) find that the
absolute value of metallicity as well as the gradient vary
over a wide range between different galaxies, which are found to correlate with
the Hubble type, luminosity and the presence or the absence of a bar.
Abundance gradients in 13 galaxies of our sample are available in 
Zaritsky et al. (1994), which are listed in Table~1. For the rest of the 
galaxies we assume a gradient of $0.05$~dex/kpc normalized to the 
solar metallicity at a radial distance of 10~kpc from the center. These
assumed values closely match the observed values for NGC\,628, 2903, 5055 
and 6946. Dust heating models by Casey~(1991) reveal that \tauvsixty\  
increases with increasing effective temperature of the heating stars and
decreasing effective grain size. Effective temperatures of stars of a
given mass are expected to be higher at the low metallicity found in the 
outer galaxies (Cervi\~no \& Mas-Hesse 1994). The low radiation 
field intensity at the outer radii favors the survival of VSGs, thus resulting
in the effective decrease of grain size with radius; the observed \Td\
gradients also support the presence of significant amount of VSGs at
the outer radius. Thus we expect an increase in \tauvsixty\  
with radius. However because of the uncertainties involved in expressing
this gradient, we prefer to fix it at a given value for each galaxy 
and comment on its effect on the observed $\tau_{60}$ gradient. 

The $\tau_{60}$ profiles are obtained for all the galaxies, based on the 
observed gas density profiles and Eq.~5, and are plotted in Fig.~5. The 
solid and dotted lines correspond to the computed \tausixty\ for 
\Sigh$=$\Sighone$+$\Sightwo\ and \Sigh$=$\Sighone\ respectively, with $b=1$. 
The effect of neglecting the dependence of gas-to-dust ratio on $Z$, i.e. 
$b=0$ in Eq.~5, is shown by the dashed line. There is one panel for each 
galaxy, with an additional panel at the end explaining the different line 
types used. The values of \tauvsixty\  are chosen such that the computed 
\tausixty\ profile matches the observed profile shape, over the maximum radial 
extent. The resulting values are indicated on the plots. It should be noted 
that the absolute values of \tauvsixty\ are based on the assumption of a 
constant gas-to-dust ratio from one galaxy to another. The observed data 
can be equally well matched by fixing \tauvsixty\ at the galactic value and 
allowing the gas-to-dust ratio to vary from galaxy to galaxy. 
From an inspection of Fig.~5, we can make the following general comments. In 
most galaxies, the \tausixty\ profile computed from \htwo\ or total gas rises
as one goes to the center, whereas the profile computed from \hi\ alone even 
with Z dependence is much flatter. The observed profile rises towards the 
center in only 8 galaxies. The observed profiles more often match the
computed profile for either \hi\ with Z dependence or the total gas with no
Z dependence which is flatter. If we consider only the sample of 13 galaxies
for which VSG corrections are available (filled circles in Fig.~5), none shows
a match with total gas with Z dependence $b=1$; four each match with only \hi\
with Z dependence and total gas with no Z dependence. In 3 galaxies no gas 
profile can give a reasonable fit.

\subsection{Spatial variation of gas-to-dust mass ratio} 

In the previous subsection we investigated the various factors responsible
for the variation of the 60\micron\ optical depth in individual galaxies.
For a constant gas-to-dust ratio in galaxies, the observed range
of \tausixty\ implies \tauvsixty\ variations over a factor of ten in different 
galaxies. This factor can be marginally reduced by invoking 
the gas-to-dust ratio to vary inversely with metallicity. However it is
conventional to derive dust masses in galaxies by fixing the values of
\tauvsixty\ corresponding to the dust properties in the solar neighborhood. 
For a 
gas-to-dust mass ratio of 160, Eq.~2 and 3 imply a value of \tauvsixty$=600$. 

We derive the profiles of the gas-to-dust ratio in galaxies by dividing the
observed profiles of gas surface density by the dust surface density. 
The resulting variation of the gas-to-dust ratio with radial distance from the 
center, expressed in units of optical radius, is shown in Fig.~6. 
Distributions of \Sighone/\Sigdust, \Sightwo/\Sigdust\ and 
 (\Sighone$+$\Sightwo)/\Sigdust\ are shown separately. At each radius there 
is one open circle for each galaxy. The sampled radii correspond to 10, 25, 
50, 75 and 100\% of the disk radius. The median values are shown joined by a
solid line. The dotted line is the median line after applying the 
estimated correction factors listed in Sec.~3.3 for the subtraction of 
contributions from VSGs and PAHs. With the adopted corrections, the dust mass 
increases by factors of 1.2, 2.3, 6.3 and 17.8 at 25, 50, 75 and 100\% of 
the disk radius respectively. In the inner regions the total gas mass 
is predominantly molecular and in the outer regions, it is mostly \hi.
The corrected radial profile of the gas-to-dust ratio is almost flat 
for the \hi\ alone, and decreases with radius for the \htwo\ and total gas. 
For the total gas profile, the gas-to-dust ratio at R$=$R$_{25}$ is 
about 215, very close to the local Milky Way value, but increases by a 
factor of 8 at the center. This smooth gradient is unlikely to represent
the real gas-to-dust mass variations and hence we investigate possible
sources for this large scale gradient. 

We first consider the effects of metallicity. It is conventional to derive 
the column density of molecular hydrogen from the CO line intensity using a
factor, $X=2.8\times10^{20}$~cm$^{-2}$/(K~km~s$^{-1}$), independent of 
metallicity of the region (Young et al. 1989). Attempts to estimate $X$ 
in external galaxies have revealed $X$ to be dependent on metallicity.  For 
example the best estimated values of $X$ in the low metallicity systems the
Large and Small Magellanic Clouds are respectively 6 and 20 times larger 
than the usually assumed value (Cohen et al. 1988; Rubin et al. 1991). 
Nakai \& Kuno (1995) also find higher values of $X$ 
in the lower metallicity outer region of M\,51. Arimoto et al. (1996) 
have compiled the estimates of the
conversion factor at different metallicities using data from several galaxies
and find that the conversion factor $X$ is inversely proportional to $Z$. 
Following this most recent study, we take  
\begin{equation}
{\rm N}_{{\rm H}_2}{\rm (true)} = {\rm N}_{{\rm H}_2}{\rm (CO)} z^{-a}, 
\end{equation}
where $z=Z/Z\odot$. The dependence of the gas-to-dust ratio with 
metallicity can also be taken as a power law, $z^b$.
Taking $a=1$ (Arimoto et al. 1996), we then consider two extreme cases
$b=1$ and $b=0$. We then expect for $b=1$, 
\begin{equation}
 {z\Sigma_{{\rm H}\,{\sc i}}+\Sigma_{{\rm H}_2}{(\rm CO})\over
{\Sigma_{\rm d}}} = {\rm constant},
\end{equation}
and for $b=0$,   
\begin{equation}
 {\Sigma_{{\rm H}\,{\sc i}}+\Sigma_{{\rm H}_2}{(\rm CO})/z\over
{\Sigma_{\rm d}}} = {\rm constant}.
\end{equation}

In Fig.~7, we plot the gas-to-dust ratio for the individual gas components
as well as the total, for both cases, $b=1$ (panels to the left), and $b=0$. 
The points and the solid lines show the values in individual galaxies and 
their median at the 5 selected radii for the 13 galaxies with metallicity 
data from Zaritsky et al. (1994). Dotted lines represent the median values 
after correction for the contamination from VSG$+$PAH. It can be seen that 
the corrected gas-to-dust ratio remains much flatter all along the
disk if the conversion factor $X$ varies inversely with $z$ (i.e. $a=1$; 
Arimoto et al. 1996), and the dust fraction does not depend on metallicity 
($b=0$). For this curve, the corrected gas-to-dust ratio beyond half the
disk radius is around 300, increasing to only about 650 at the center.

The derived absolute value of the gas-to-dust ratio depends on many
uncertainties like the absolute value of $X$, its dependence with $z$,
the FIR dust emissivity and the contribution of VSG and PAH to the FIR 
emission and the resultant correction to \tausixty\ etc. 
In the present study we have shown that the gradient in the gas-to-dust
ratio is very sensitive to the corrections to all the above effects. 
With our most realistic corrections, the gas-to-dust ratio in the outer
galaxies turns out to be not very much different from the local Galactic 
value of 160. However, the central value is still higher by factor of 3--4.
This can arise from the fact that while in the outer regions, the gas is 
predominantly diffuse \hi\ and is less dense, in the inner region, 
it is molecular, which is denser and more clumpy. Not all the dust associated 
with the molecular gas would then be heated to the warm temperature 
detectable by IRAS. Recent studies of normal ``in-active'' galaxies using 
mm continua indicate temperatures ranging from 10--20~K for the cold 
dust (Chini et al.  1995), consistent with our results.
Sub-mm observations with the Infrared Space Observatory (ISO),
with its capability for high angular resolution, will have a major role 
to play in elucidating the role of cool dust and its location in
galaxies.

\subsection{Global gas-to-dust ratio}        

Young et al. (1989) investigated correlations between the global dust mass
and neutral and molecular hydrogen, and came to the conclusion that
the IRAS-detected dust is closely associated with the molecular gas. 
The radial profiles of the optical depth of our sample galaxies are
systematically flatter than that of \htwo, implying that the dust is not 
preferentially associated with the molecular gas locally within galaxies.
We carry out the following analysis to sort out this apparent discrepancy. 
Cumulative dust and gas masses in our
program galaxies are computed by summing over the radial profiles of dust
and gas distribution. The resultant gas-to-dust mass ratios are plotted in
Fig.~8 as a function of the fractional disk radius (3 panels on the left). 
The open circles and the solid lines have the same meaning as in Fig.~6. 
The dotted lines represent 
cumulative values after using the corrected dust and recomputed 
molecular ($a=1$) masses. We use $b=0$ following the results from Fig.~7.
Only 13 galaxies with the metallicity data are used in these plots.
From this figure it is noted that the dispersion close to the center is 
larger for both \hi\ and \htwo\ than in the outer (global) regions. 
For the global value (R/R$_{25} = 1$), the dispersion is again similar for 
\hi\ and \htwo. Thus we find that the dust mass is equally well correlated
with both the \hi\ and \htwo\ gas for this small sample. The presence of 
a non-negligible \hi\ mass outside the optical radius may be the reason for 
the relatively poor correlation between global dust and \hi\ masses
as seen by Young et al. (1989).

The global value of 300 for the gas-to-dust ratio is, within a factor of
two, the same as the canonical value. Note that the global value
is representative of the outer disk rather than the higher
intensity inner disk value of 600, contrary to the general belief. Why
it is so can be understood from the three panels on the right in Fig.~8. 
Cumulative masses of the total gas, molecular hydrogen and dust are plotted
separately normalized by the respective global masses (at R$=$R$_{25}$). 
As in the previous plots open circles and solid lines show the observed
masses, while the dotted lines show the corrected masses. The panel for 
the dust mass also includes the cumulative FIR luminosity (defined by the 
numerator of Eq.~1), again normalized by the total FIR luminosity. These 
plots reveal that the gas and dust lying outside half the disk radius 
contribute as much as 60\% to the total gaseous mass and dust 
mass. Thus outer regions control the global gas-to-dust ratio in galaxies.

The availability of radial profiles of dust and gas mass distributions
gives us an opportunity to comment on the wide range of global gas-to-dust 
mass ratios reported in the literature. The most important results from
this study on the gas and dust mass distributions are summarized in Table~2,
to ease comparisons with other data sets. The corrected dust mass
that we compute, based on the D\'esert et al. (1990) model, takes into account 
all the mass in dust particles hotter than 20~K. The dust 
masses derived from the global IRAS data are weighted by the 
FIR intensity, 90\% of which comes from within half
the disk radius, and hence pertain only to the central value. If we follow the
usual practice of neglecting the $Z$-dependence of the CO intensity to \htwo\ 
mass conversion factor, we derive a value between 700--1350 (Col.~9), 
for the global gas-to-dust mass ratio. This explains the results from Devereux
 \& Young (1990), who obtained a value of around 1000 using global IRAS data.
Correcting the molecular masses for the effects of $Z$ decreases the total gas
mass by only around 20\% (Col.~5), but drastically flattens the radial 
distribution of the gas-to-dust ratio (Cols.~7 and 8). Thus without these 
corrections, observed
gas-to-dust ratios are expected to be highly aperture dependent even if
the dust masses are determined using mm and sub-mm aperture observations,
as illustrated in Col.~10. Andreani et al. (1995), using mm dust continuum 
observations and a $X$ value reduced by 1.6 times, derive a global gas-to-dust 
ratio of 230, in good agreement with our corrected global value (Col.~11).

\section{Conclusions}

Gradients in the warm dust temperature within spiral galaxies are studied
for a sample of 22 galaxies with low levels of present star formation.
The observed \Td\ is found to decrease away from the center, followed by
a slow increase in the outer parts of galaxies. 
We invoked the D\'esert et al. (1990) models to estimate the heating of dust
by non-ionizing photons from the bulge and disk stars in 13 of 
these galaxies. The model reproduces the observed \Td\ gradients
qualitatively well, with very small grains playing a major role in maintaining 
high \Td\ values at regions of low flux levels such as in the outer 
regions of galaxies. However, additional heating sources from recent star 
forming events are required for a detailed matching of the \Td\ profiles.
Correction for the contribution from very small grains is essential to 
derive the optical
depth due to the big grains alone, which dominate the dust mass.
Surface density profiles of dust are flatter than that of molecular hydrogen 
in nearly 50\% of the galaxies, implying that only a part of the dust mass
associated with the molecular hydrogen contributes to the FIR emission.
We further showed that the often reported high values ($\sim1000$) for 
the gas-to-dust ratios in galaxies from global IRAS data, are due to a
combination of over-estimation of molecular masses (as a result of using
a metallicity-independent conversion factor from CO intensity to \htwo\ column
density), and due to the presence of significant amounts of cool dust in 
the outer fainter parts of galaxies. When corrections are made for these
effects, radial profiles of the gas-to-dust mass ratio flatten out, 
and the global value approaches, to within a factor of two, the
local galactic value.

\begin{acknowledgements}

 It is a pleasure to thank Dr. Walter Rice, the referee, for his comments 
 and suggestions towards improvement of the manuscript.
 We wish to thank Dr. Z.\,Frei and collaborators for making their reduced 
 CCD data of galaxies available for public use.
 This research has made use of the NASA/IPAC Extragalactic Database (NED)   
 which is operated by the Jet Propulsion Laboratory, California Institute   
 of Technology, under contract with the National Aeronautics and Space      
 Administration.                                                            
  
\end{acknowledgements}

\clearpage

\onecolumn


\newcommand\nod{\nodata}

\begin{deluxetable}{llrrccrcrcccr}
\tablenum{1}
\tablewidth{0pt}
\tablecaption{Basic data on sample galaxies}
\tablehead{
\colhead{NGC}           & \colhead{Type}      &  
\colhead{dist}          & \colhead{S$_{60}$}  &  
\colhead{L$_{\rm fir}$/L$_{\rm B}$}           &  
\colhead{$60\mu$m beam} & \colhead{D$_{25}^\prime$}  &  
\colhead{b/a}           & \colhead{PA}        &
\colhead{$Z$\tablenotemark{1}} & 
\colhead{Z-grad}        &  \multicolumn{2}{c}{Ref} \\  
\colhead{}              & \colhead{}          &
\colhead{Mpc}           & \colhead{Jy}        &  
\colhead{}              & 
\colhead{$^{\prime\prime}$ x $ {^{\prime\prime}}$} & 
\colhead{}              & \colhead{}          &  
\colhead{$^\circ$}       & \colhead{} &      
\colhead{dex/kpc}       & \colhead{H\,{\sc i}} &  \colhead{H$_2$} }  
\startdata
 628 & SA(s)c    &  9.7 & 25.5 & 3.41 &  77 x 42 & 10.47 & 0.91 &  25 & 9.13 & $-$0.063 & 1 & 9  \\
2403 & SAB(s)d   &  4.2 & 62.6 & 2.04 &  60 x 42 & 21.88 & 0.56 & 127 & 8.80 & $-$0.058 & 1 & 9 \\
2841 & SA(r)b    & 12.0 &  6.3 & 1.01 &  79 x 45 &  8.13 & 0.44 & 147 & \nod & \nod     & 2 & 9 \\
2903 & SAB(rs)bc &  6.3 & 67.6 & 3.70 &  68 x 45 & 12.59 & 0.48 &  17 & 9.22 & $-$0.058 & 1 & 9 \\
3079 & SB(s)c    & 20.4 & 52.8 & 9.90 &  40 x 35 &  7.94 & 0.18 & 165 & \nod & \nod     & 3 & 9 \\
3198 & SB(rs)c   & 10.8 & 6.9  & 1.43 &  67 x 32 &  8.51 & 0.39 &  35 & 8.94 & $-$0.065 & 1 & \nod \\
3627 & SAB(s)b   &  6.6 & 66.5 & 4.22 &  72 x 42 &  9.12 & 0.46 & 173 & \nod & \nod     & 4 & 9 \\
3628 & ScP       &  7.7 & 56.9 & 4.17 &  86 x 35 & 14.79 & 0.20 & 104 & \nod & \nod     & 5 & 9 \\
4192 & SAB(s)ab  & 16.8 &  8.4 & 1.40 &  92 x 45 &  9.77 & 0.28 & 155 & \nod & \nod     & 6 & 10 \\
4254 & SA(s)c    & 16.8 & 40.7 & 6.41 &  64 x 43 &  5.37 & 0.87 &   0 & 9.26 & $-$0.035 & 6 & 10 \\
4258 & SAB(s)bc  &  6.8 & 27.4 & 1.16 &  63 x 43 & 18.62 & 0.87 & 150 & 9.07 & $-$0.031 & 1 & 9 \\
4303 & SAB(rs)bc & 15.2 & 40.2 & 6.01 &  83 x 45 &  6.46 & 0.89 &   0 & 9.20 & $-$0.078 & 6 & 10 \\
4321 & SAB(s)bc  & 16.8 & 26.9 & 4.10 &  55 x 35 &  7.41 & 0.85 &  30 & 9.32 & $-$0.012 & 6 & 10 \\
4501 & SA(rs)b   & 16.8 & 20.7 & 3.32 &  61 x 44 &  6.92 & 0.54 & 140 & \nod & \nod     & 6 & 10 \\
4535 & SAB(s)c   & 16.8 & 12.6 & 2.52 &  62 x 44 &  7.08 & 0.71 &   0 & \nod & \nod     & 6 & 10 \\
4569 & SAB(rs)ab & 16.8 & 11.0 & 1.33 &  84 x 44 &  9.55 & 0.46 &  23 & \nod & \nod     & 6 & 10 \\
4736 & RSA(r)ab  &  4.3 & 75.2 & 2.76 &  63 x 42 & 11.22 & 0.81 & 105 & 9.00 & $-$0.025 & 7 & 9 \\
5033 & SA(s)c    & 18.7 & 21.6 & 3.99 &  65 x 43 & 10.72 & 0.47 & 170 & 9.09 & $-$0.035 & 1 & 9 \\
5055 & SA(rs)bc  &  7.2 & 50.9 & 3.95 &  61 x 42 & 12.59 & 0.58 & 105 & 9.31 & $-$0.058 & 1 & 9 \\
6503 & SA(s)cd   &  6.1 & 11.1 & 2.09 &  67 x 44 &  7.08 & 0.34 & 123 & \nod & \nod     & 1 & 9 \\
6946 & SAB(rs)cd &  5.5 &165.2 & 2.87 &  39 x 35 & 11.48 & 0.85 &  32 & 9.13 & $-$0.052 & 8 & 9 \\
7331 & SA(s)b    & 14.3 & 41.9 & 4.06 &  75 x 43 & 10.47 & 0.36 & 171 & 9.15 & $-$0.024 & 2 & 9 \\
\tablenotetext{1}{12$+\log$(O/H) at radius 3 kpc} 
\tablerefs{
(1) Wevers et al. 1986; (2) Begeman 1987;        (3) Irwin \& Seaquist 1991; 
(4) Zhang et al. 1993;  (5) Wilding et al. 1993; (6)~Warmels~1988;  
(7)~Bosma 1978;  
(8)~Tacconi \& Young 1986; (9) Young et al. 1995; (10) Kenney \& Young 1988 
}
\enddata
\end{deluxetable}

\begin{deluxetable}{crcccrrcrrc}
\tablenum{2}
\tablewidth{0pt}
\tablecaption{Mean properties of gas and dust masses in sample galaxies}
\tablehead{
\colhead{R/R$_{25}$}                                     & 
\colhead{$f_{\Sigma_{\rm D}}$}                           & 
\colhead{$f_{{\rm M}_{\rm D}}$}                          & 
\colhead{$f_{{\rm M}_{{\rm H}_2}}$}                      & 
\colhead{$f_{{\rm M}_{\rm H}}$}                          & 
\colhead{$\Sigma_{\rm H}/\Sigma_{\rm D}$}                & 
\colhead{$\Sigma_{\rm H}/\Sigma_{\rm D}^\prime$}         & 
\colhead{$\Sigma_{\rm H}^\prime/\Sigma_{\rm D}^\prime$}  &
\colhead{M$_{\rm H}$/M$_{\rm D}$}                        & 
\colhead{M$_{\rm H}$/M$_{\rm D}^\prime$}                 & 
\colhead{M$_{\rm H}^\prime$/M$_{\rm D}^\prime$}          \\ 
\colhead{1} & \colhead{2} & \colhead{3} & \colhead{4} & \colhead{5} & 
\colhead{6} & \colhead{7} & \colhead{8} & \colhead{9} & \colhead{10} & 
\colhead{11} } 
\startdata
0.1 &  0.90 & 1.00 & 0.30 & 0.35 & 1567 & 1761 & 655 & 1349 & 1318 & 575 \\
0.5 &  2.34 & 1.55 & 0.54 & 0.76 &  955 &  408 & 274 &  692 &  501 & 316 \\
1.0 & 17.78 & 2.75 & 0.54 & 0.79 & 3802 &  214 & 335 &  832 &  347 & 275 \\
\tablecomments{
Col.\,2: The adopted mean correction factor for dust mass at the three
chosen radii, following the discussion in sec.~3.3.\\ 
Cols.\,3--5: Ratios of the uncorrected to the corrected cumulative masses, for 
the dust, H$_2$ and the total gas respectively. Correction to H$_2$
mass involves the usage of metallicity dependent conversion factor between 
CO line intensity and H$_2$ mass. \\
Cols.\,6--8: Gas-to-dust surface mass ratios at the three sampled radii,
with primes denoting the quantities after corrections, following our
prescriptions described above.\\
Cols.\,9--11: Cumulative gas-to-dust mass ratios, with and without corrections.
As in the earlier columns, primes indicate corrected quantities.
}
\enddata 
\end{deluxetable}


\begin{figure}[htb]
\vspace*{-2.0cm}
\centerline{\psfig{figure=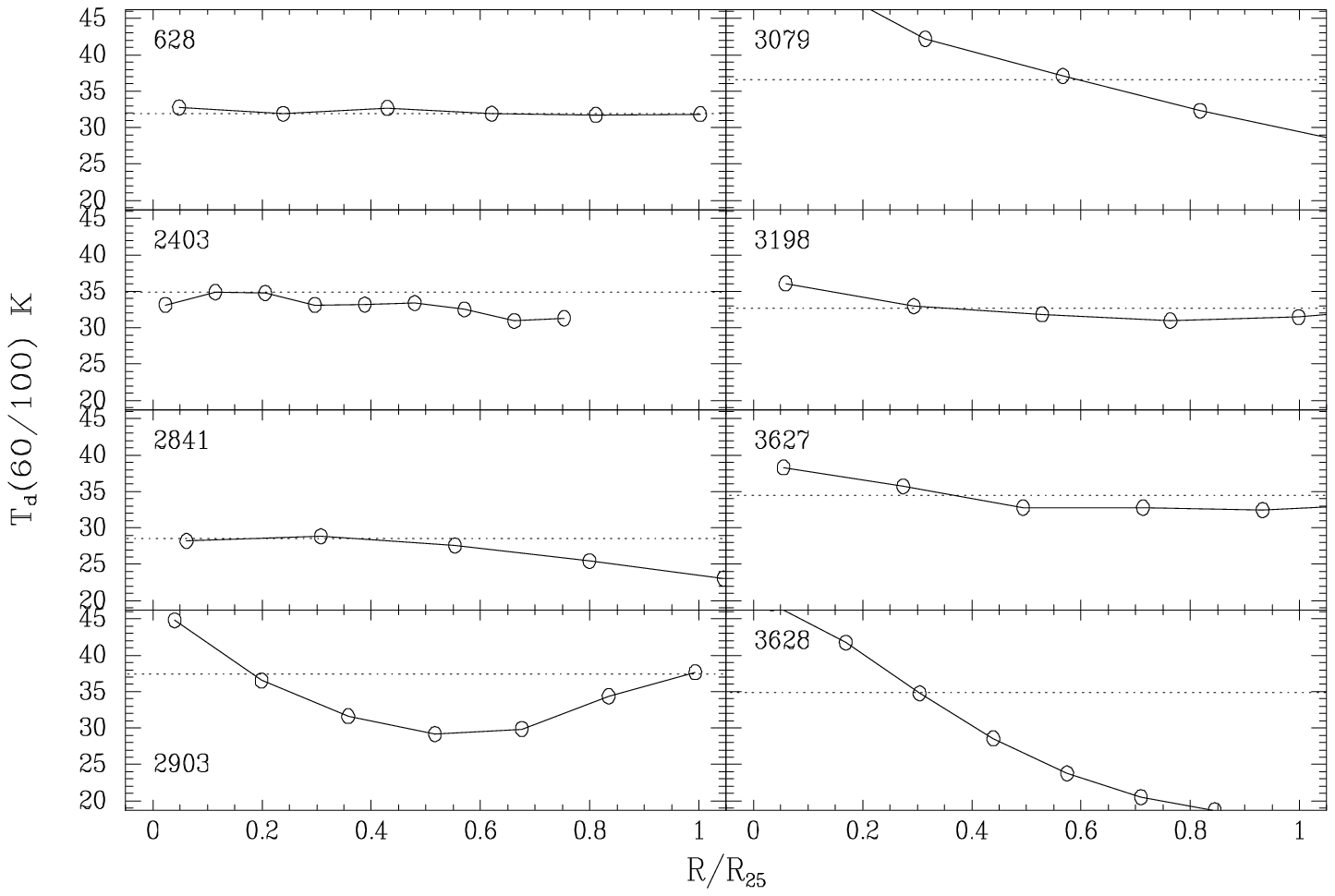,height=11.0cm}}
\vspace*{-1.0cm}
\centerline{\psfig{figure=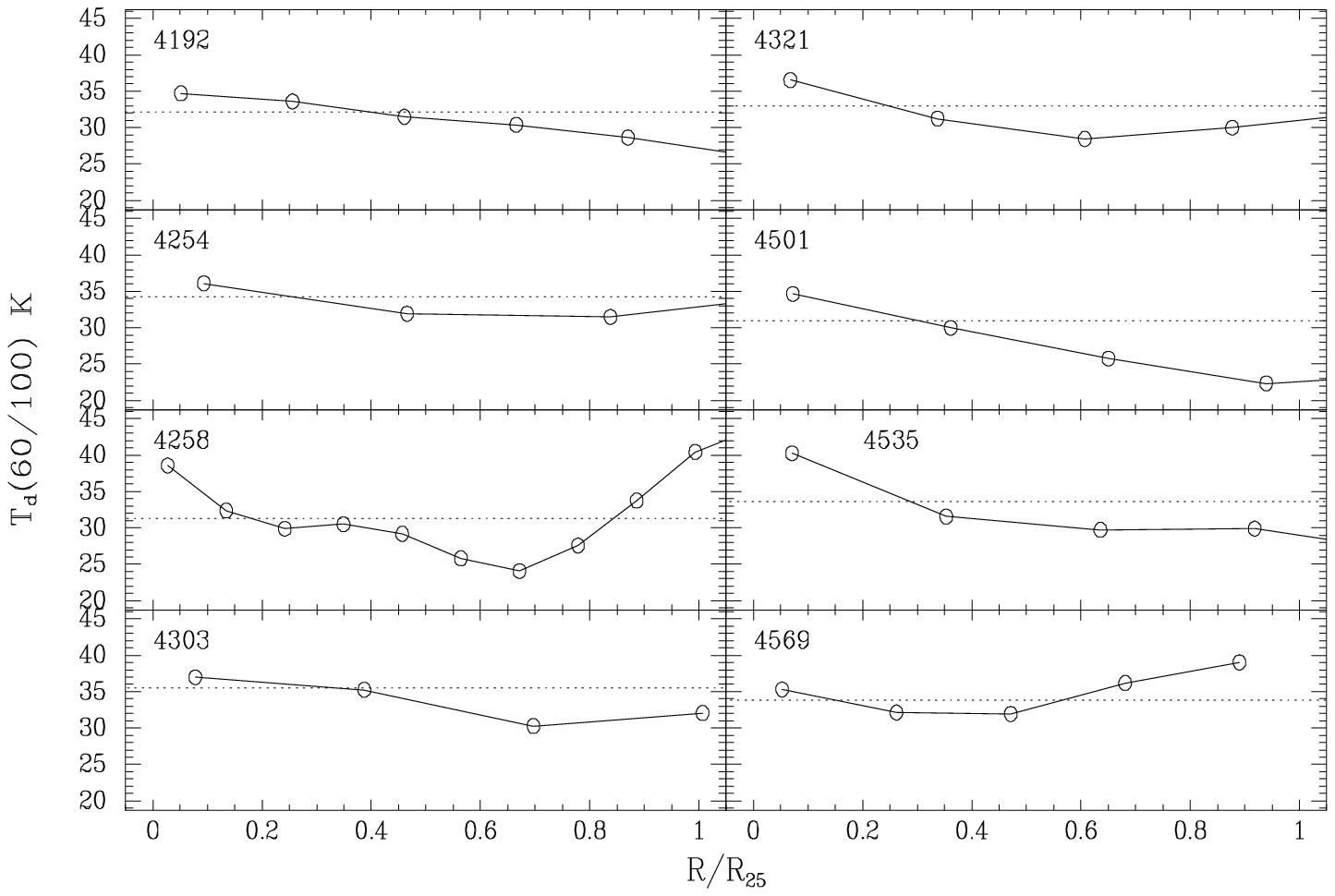,height=11.0cm}}
\vspace*{-0.5cm}
\caption{
Observed warm dust temperature profiles (open circles at 1\arcmin\ intervals
joined by the solid line) in 22 galaxies are plotted as a function of 
distance from the center, expressed in units of disk radius. 
NGC numbers of galaxies are given on the left side in each panel.
The dotted lines correspond to the \Td\ values derived from the global data.
The same range of \Td\ is plotted in all the galaxies to enable easy
comparison of gradients and absolute values of different galaxies.
\Td\ decreases away from the center increasing again by a few Kelvin at 
the outer radii in the majority of the galaxies. 
}
\end{figure}

\begin{figure}[htb]
\vspace*{-1.5cm}
\centerline{\psfig{figure=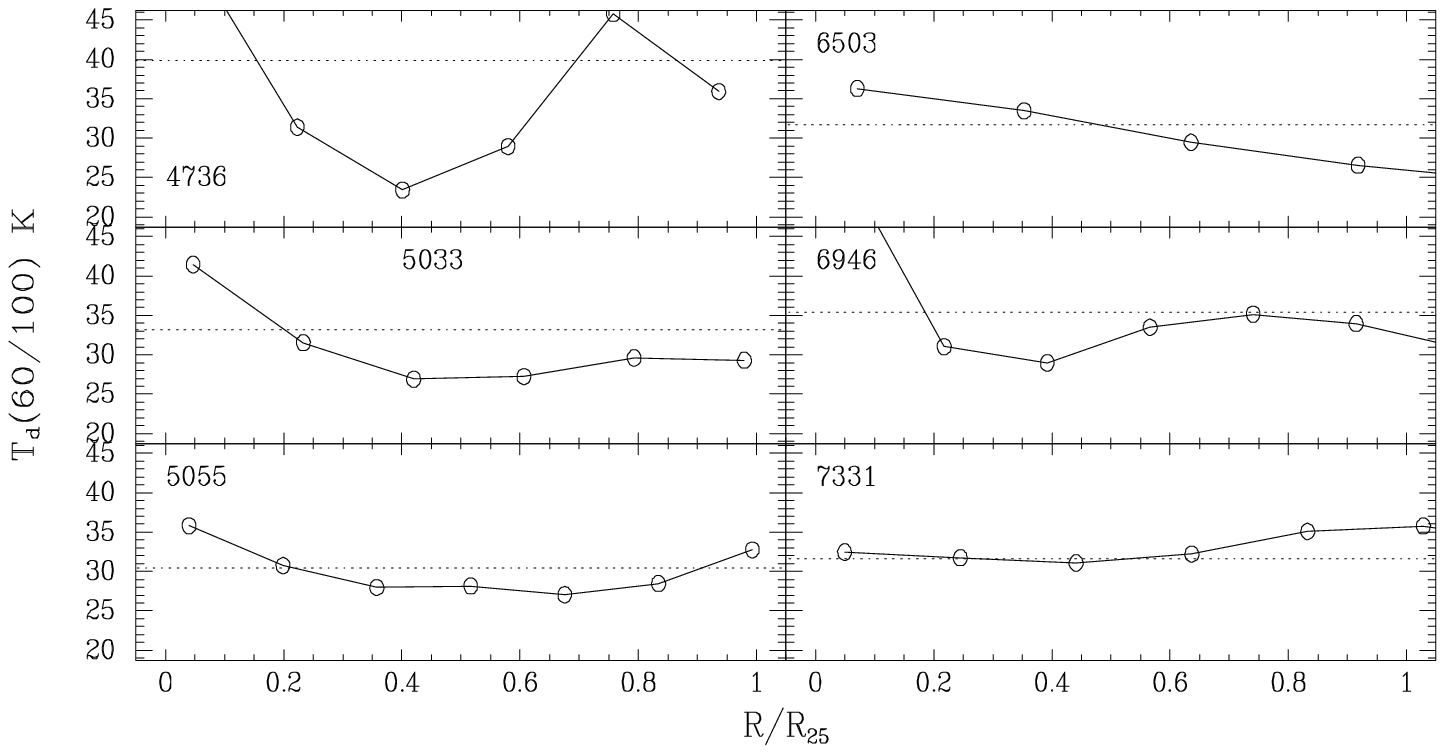,height=10.0cm}}
\vspace*{-0.5cm}
\centerline{Fig.~1.--- Continued.}

\centerline{\psfig{figure=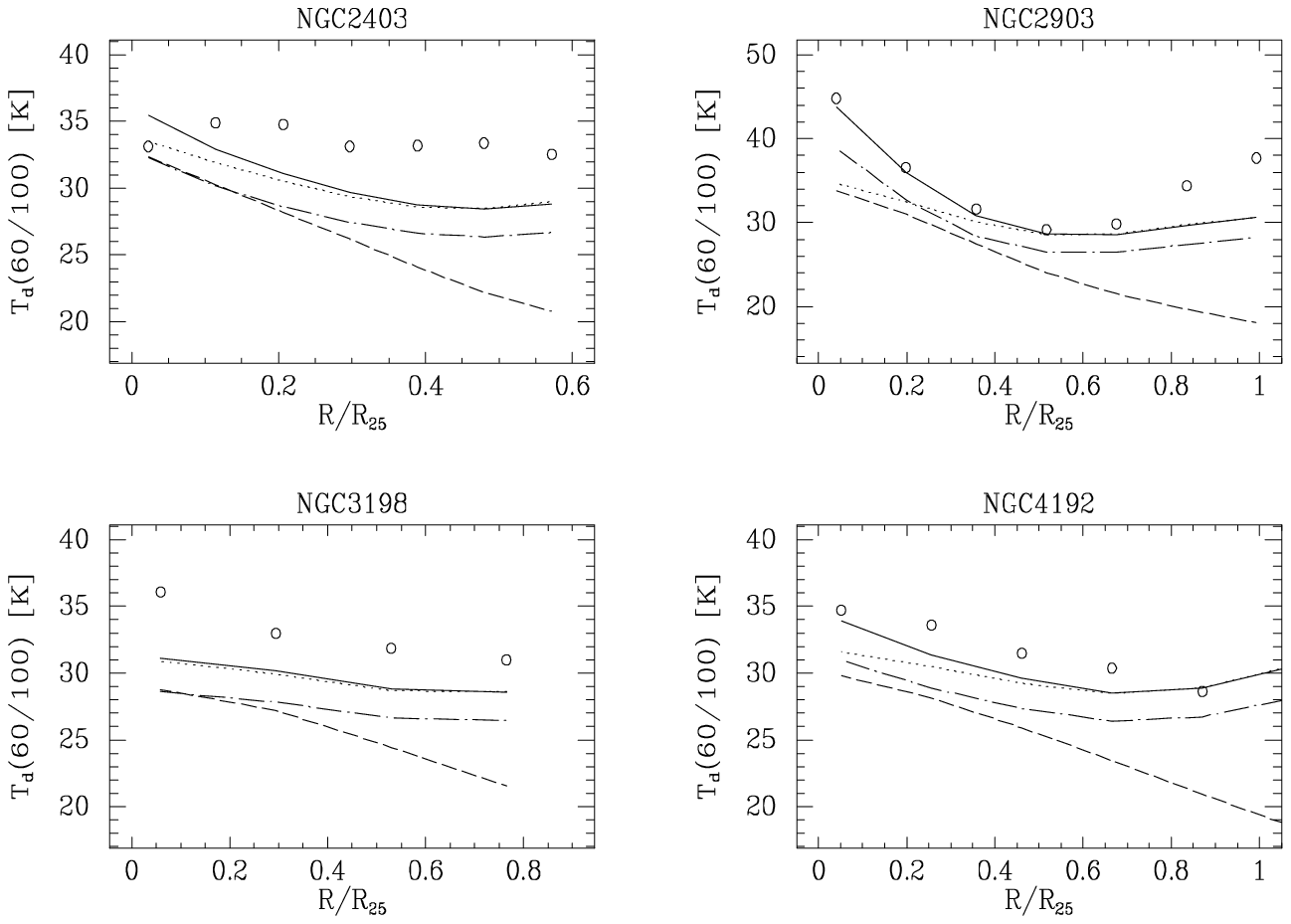,height=11.0cm}}
\caption{
Observed \Td\ gradients in 13 galaxies are compared with the dust heating
by the interstellar radiation field as modeled by D\'esert et al. (1990). 
The last panel explains the meaning of the different lines. The observed
data points are denoted by the open circles. Note the importance of small
grains (VSGs$+$PAHs) in reproducing the outward increase of \Td\ beyond
half the disk radius in majority of the galaxies. 
}
\end{figure}

\begin{figure}[htb]
\vspace*{-2.0cm}
\centerline{\psfig{figure=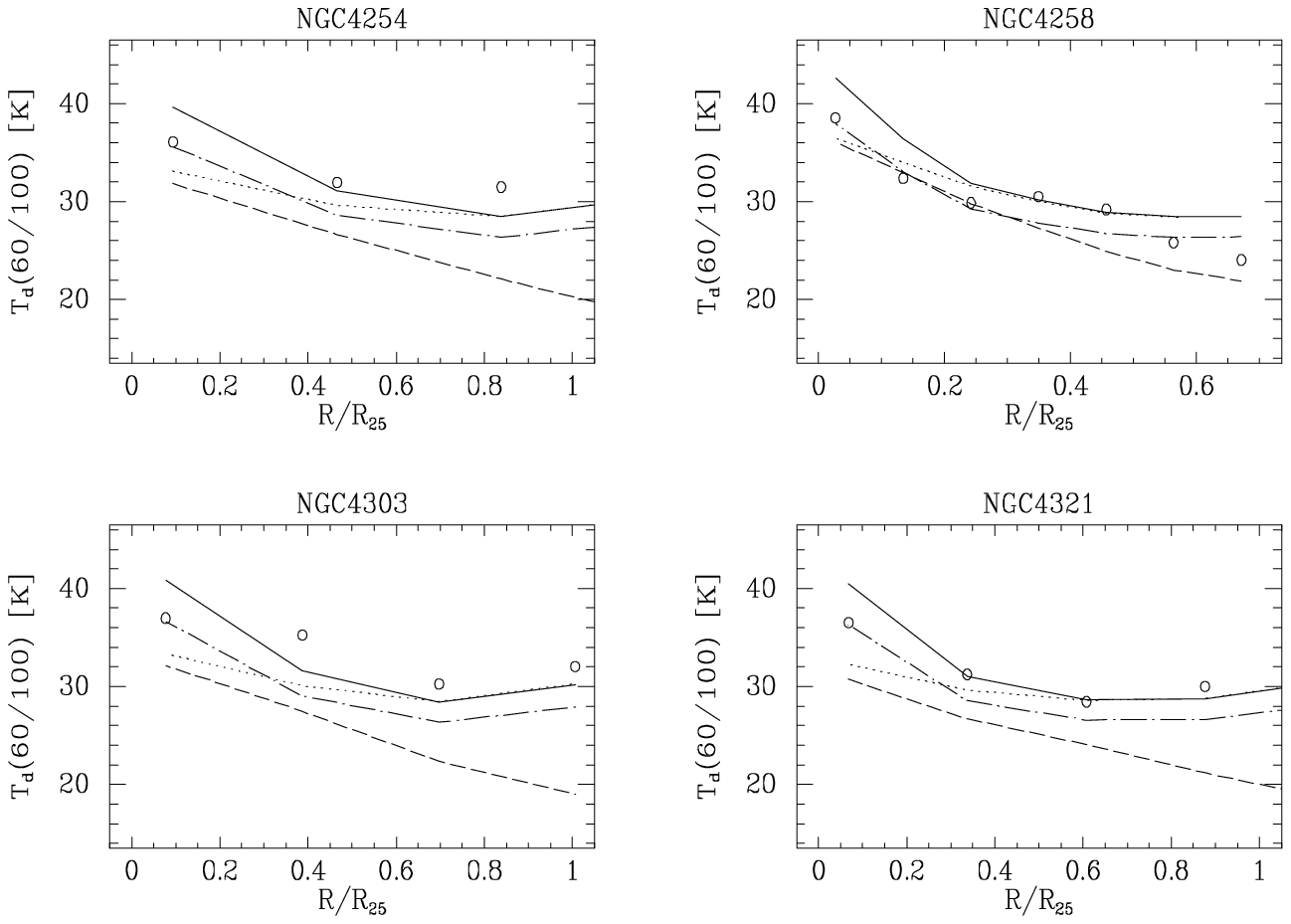,height=11.0cm}}
\vspace*{-1.0cm}
\centerline{\psfig{figure=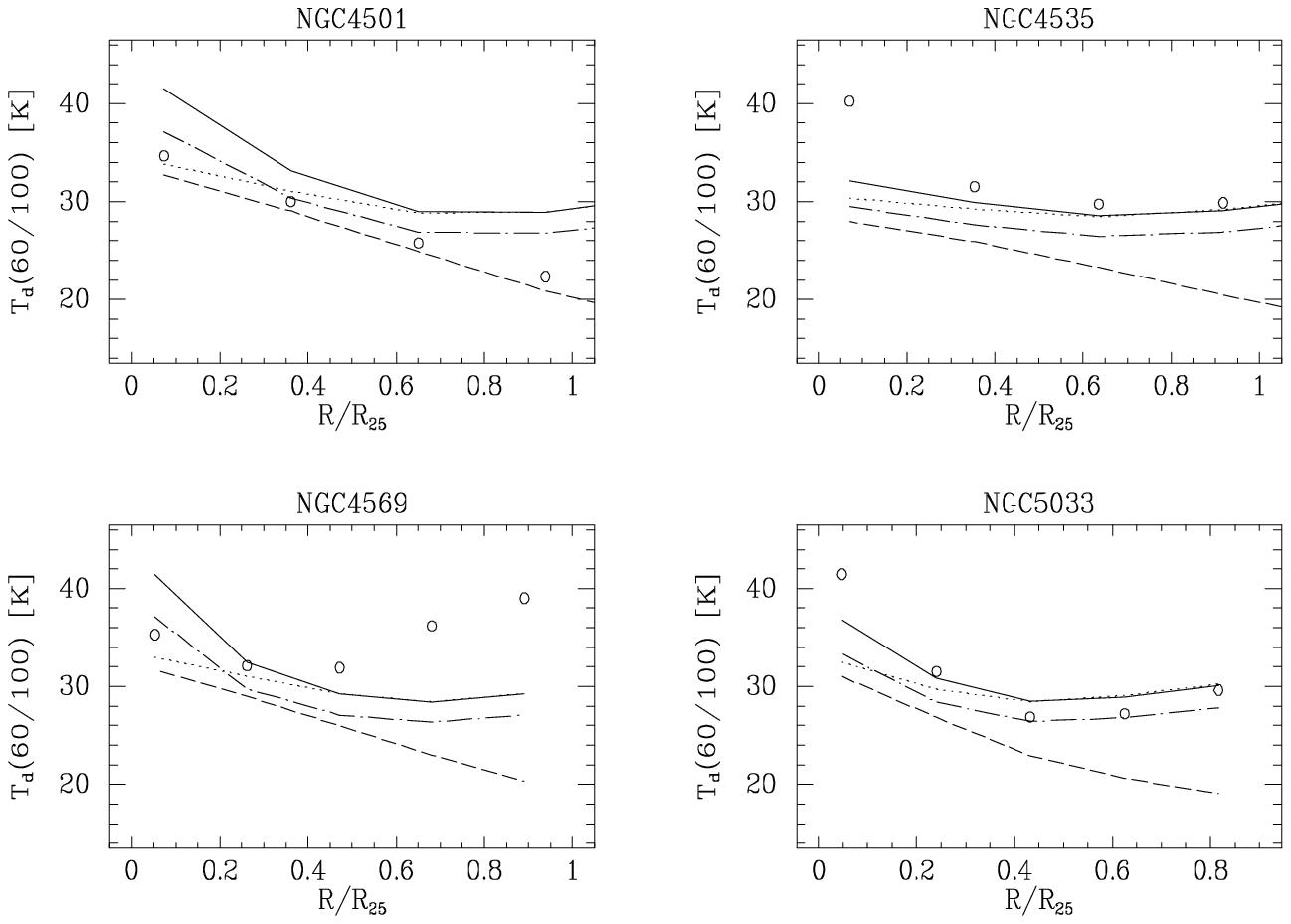,height=11.0cm}}
\vspace*{-0.5cm}
\centerline{Fig.~2.--- Continued.}
\end{figure}

\begin{figure}[htb]
\vspace*{-2.5cm}
\centerline{\psfig{figure=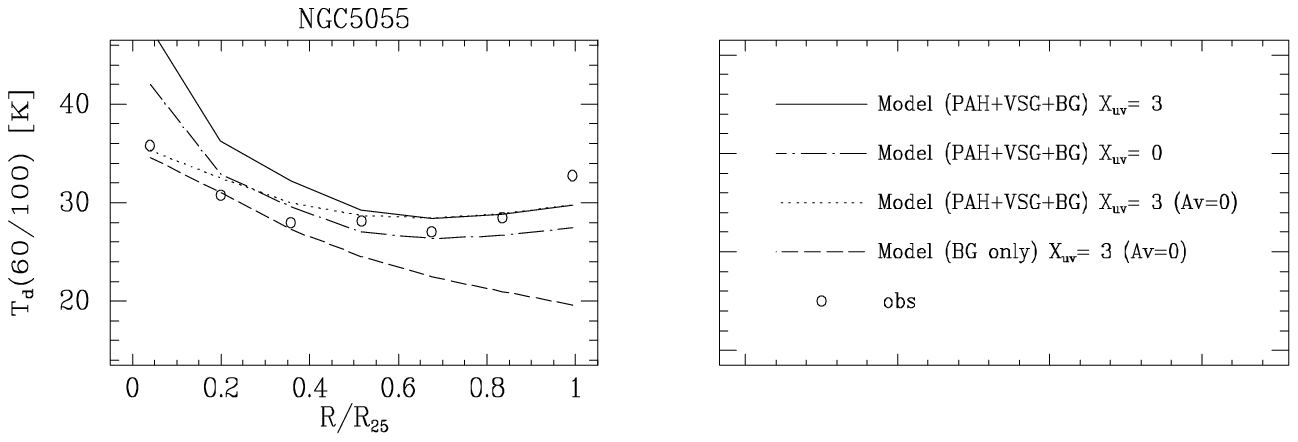,height=11.0cm}}
\vspace{-5.5cm}
\centerline{Fig.~2.--- Continued.}

\vspace{0.5cm}
\centerline{\psfig{figure=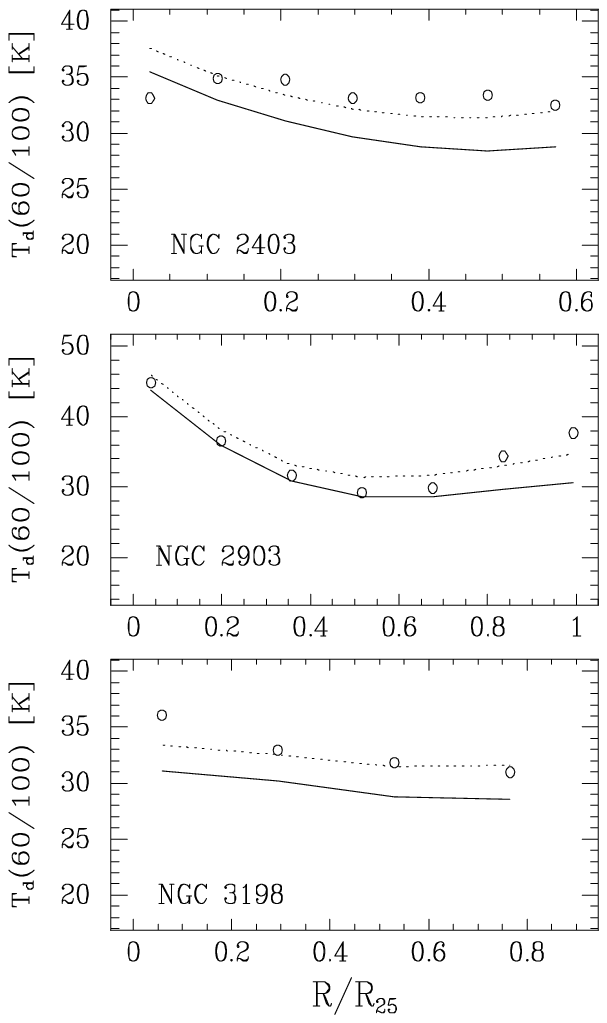,height=13.0cm}}
\vspace{-0.3cm}
\caption{
\Td\ radial gradients in three galaxies with the largest difference between the
observed and modeled gradients are plotted. Open circles and the solid 
lines are the same as plotted in Fig.~2. The dotted line corresponds to a 
model, in which the flux contribution to the 60 and 100\micron\ bands from 
VSGs are doubled with respect the values expected from the 
D\'esert et al. (1990) model. 
}
\end{figure}

\begin{figure}[htb]
\vspace*{-0.0cm}
\centerline{\psfig{figure=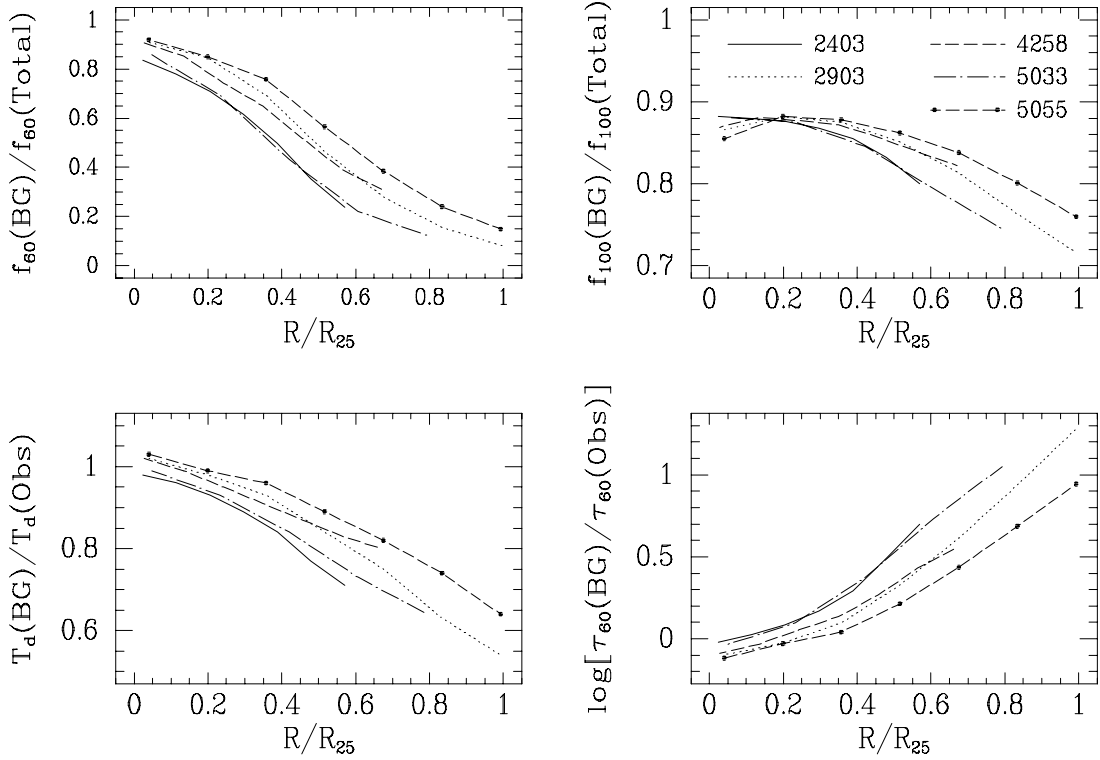,height=12.0cm}}
\vspace*{-0.0cm}
\caption{
The contribution to the 60 and 100\micron\ emission from the big grains alone, 
as fraction of the total emission, is plotted as a function of the radial 
distance for five representative galaxies in the sample in the top two
panels. The dust temperature and optical depth are also computed taking
into account the contribution from big grains alone and are plotted as 
fractions of the corresponding observed values, in the other two panels. 
The fact that all of the five galaxies show similar radial
behavior, helps in estimating the contributions from
the big grains alone in the rest of the galaxies.
}
\end{figure}

\begin{figure}[htb]
\vspace*{-2.0cm}
\centerline{\psfig{figure=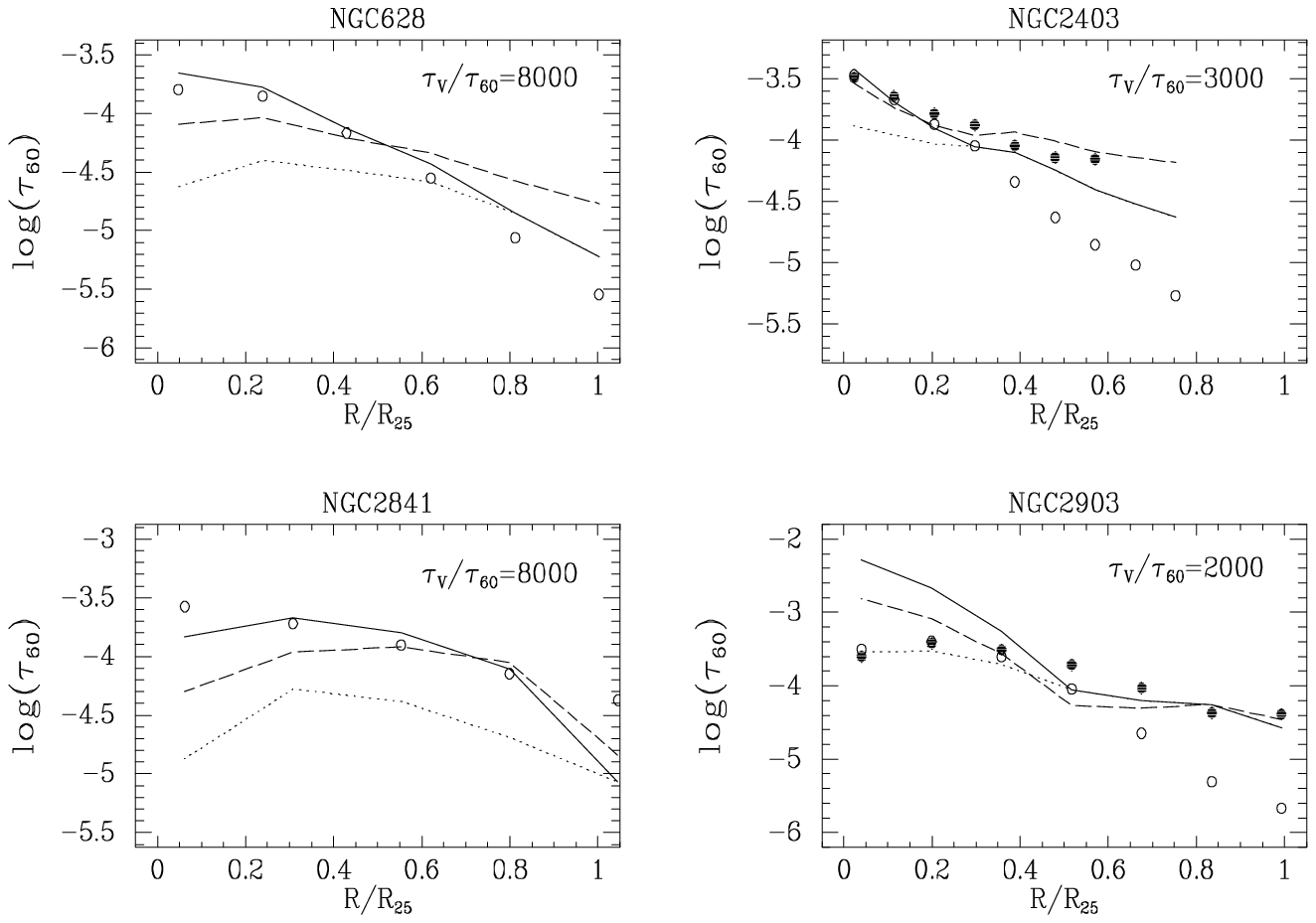,height=11.0cm}}
\vspace*{-1.0cm}
\centerline{\psfig{figure=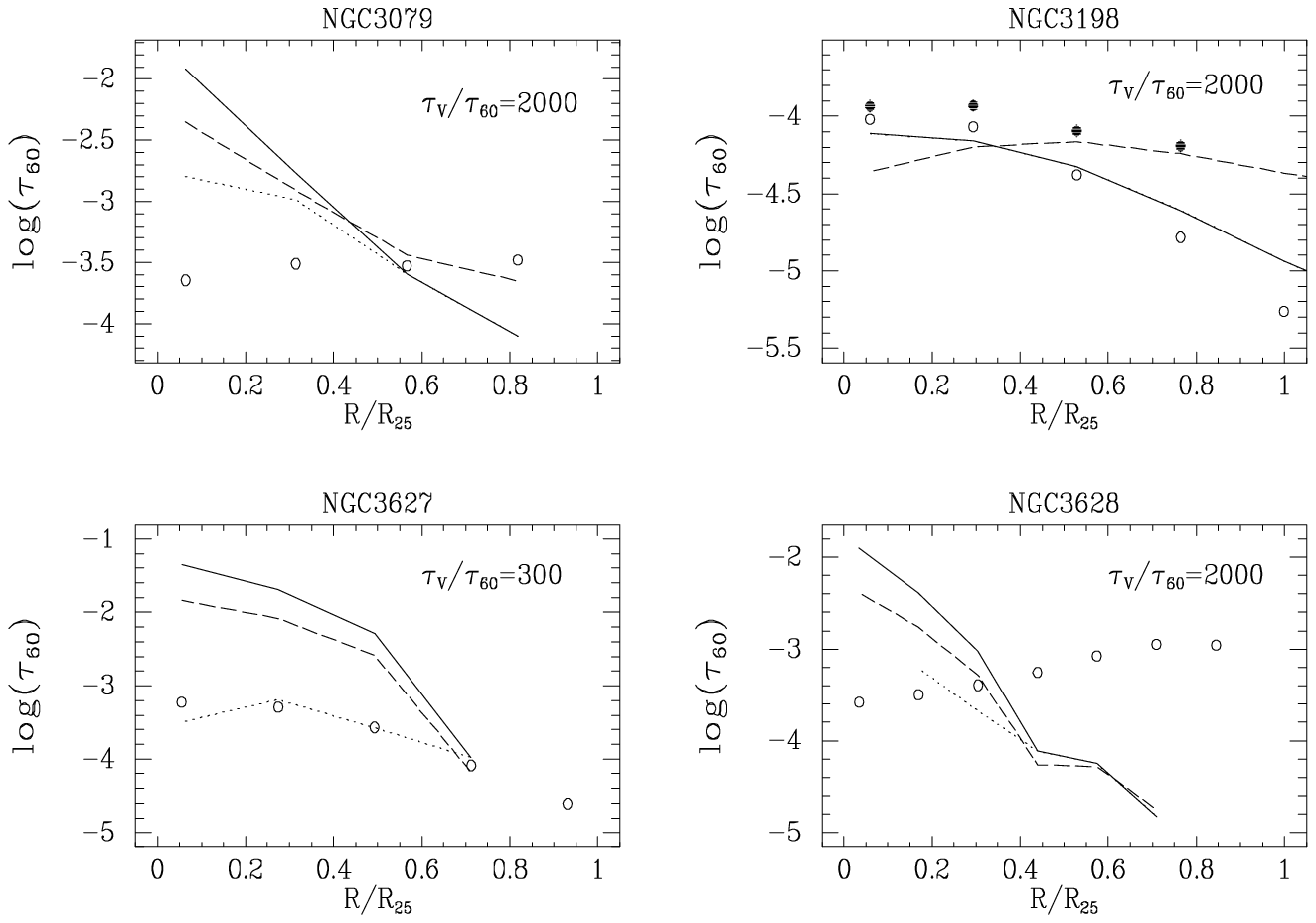,height=11.0cm}}
\vspace*{-0.5cm}
\caption{
The observed $\tau_{60}$ radial profiles are compared with that expected from 
Eq.~5 for the total and atomic hydrogen gas surface densities separately.
Open circles are the observed values, whereas the the filled circles
(for 13 galaxies) are the estimated \tausixty\ values for the big grains 
alone which make up almost the total dust mass column density.
The assumed value of \tauvsixty\ is indicated for each galaxy.
The last panel explains the meaning of the different lines. 
See Sec.~4.1 for details.
}
\end{figure}

\begin{figure}[htb]
\vspace*{-2.0cm}
\centerline{\psfig{figure=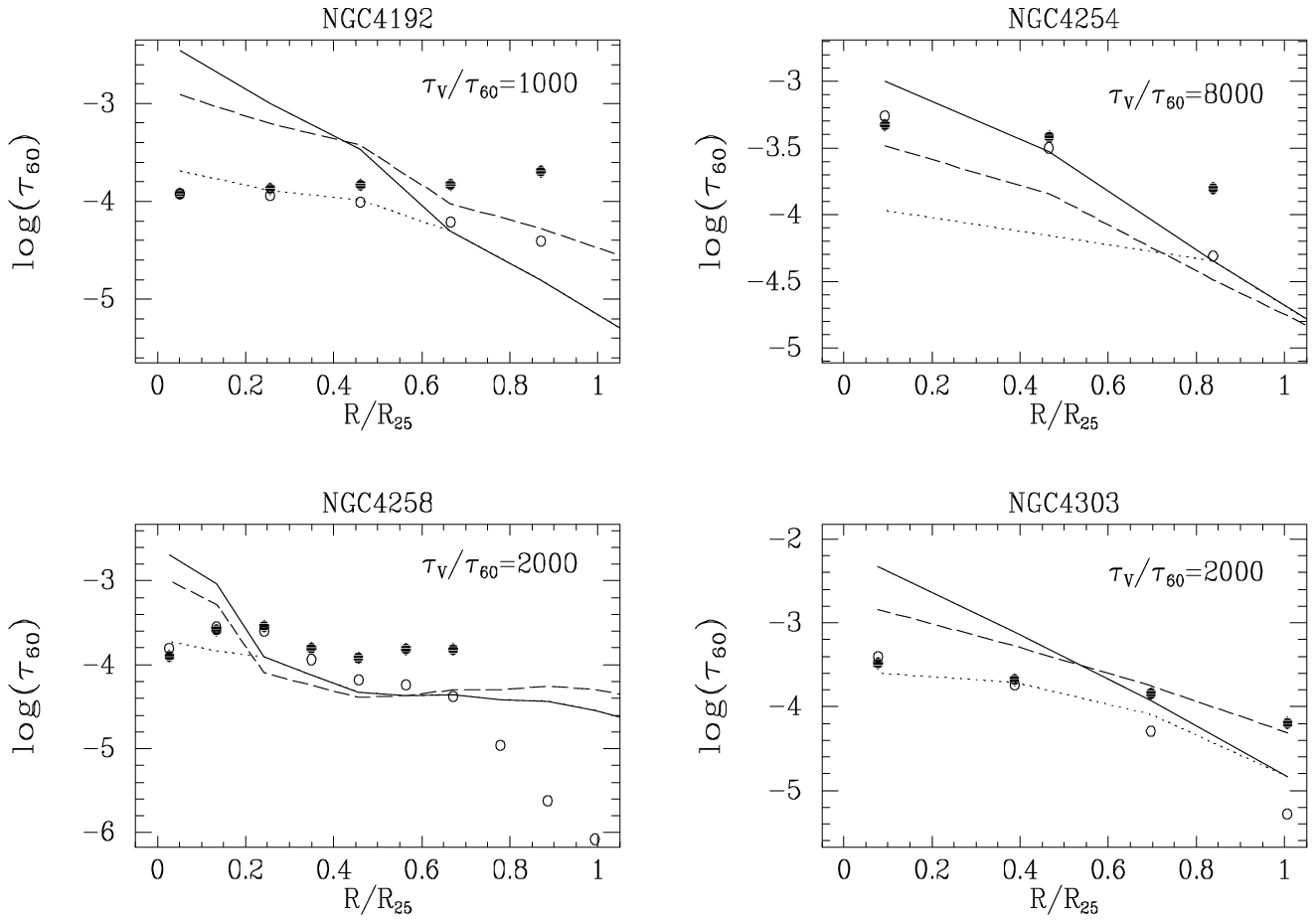,height=11.0cm}}
\vspace*{-1.0cm}
\centerline{\psfig{figure=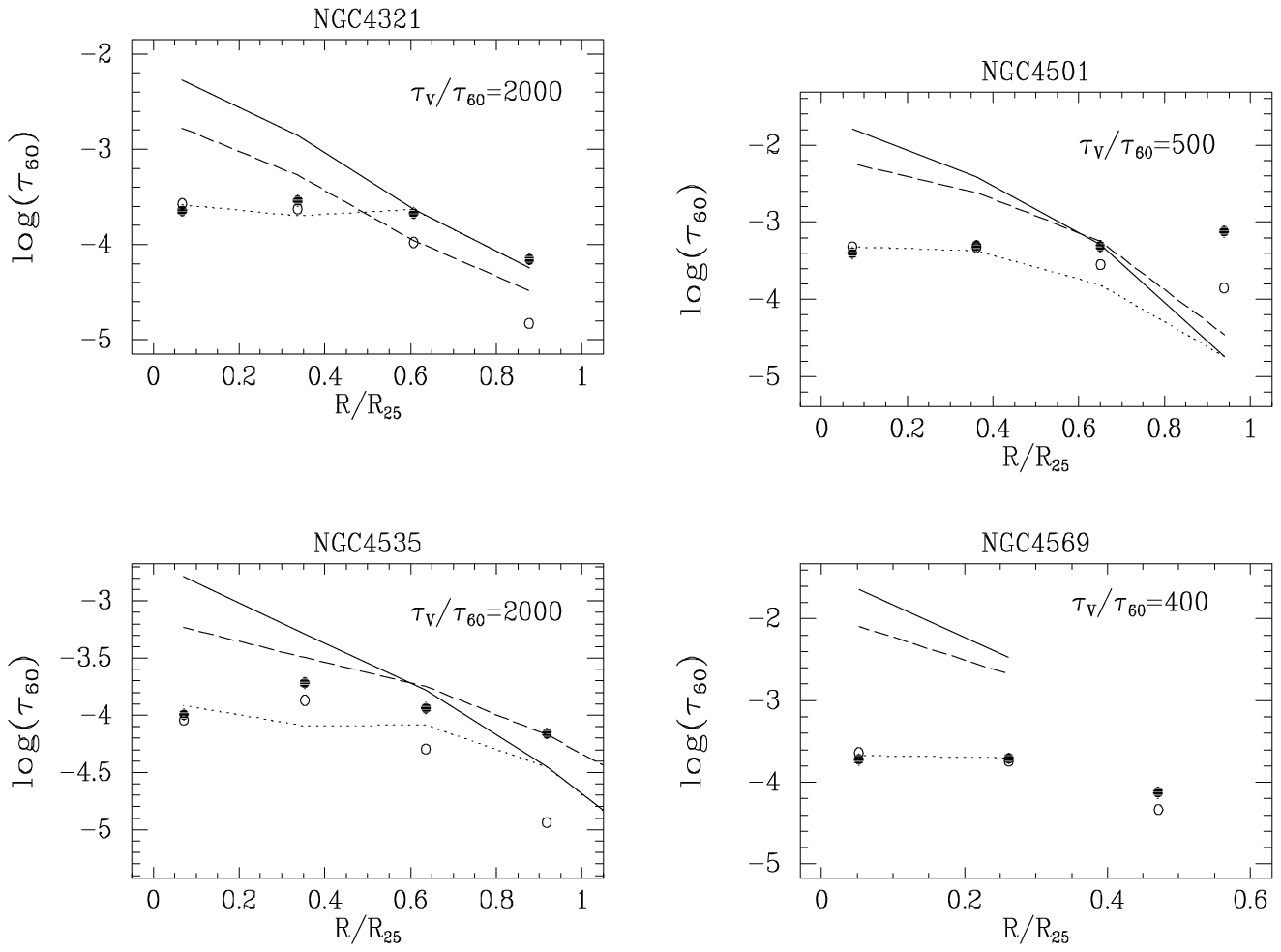,height=11.0cm}}
\vspace*{-0.5cm}
\centerline{Fig.~5.--- Continued.}
\end{figure}

\begin{figure}[htb]
\vspace*{-2.0cm}
\centerline{\psfig{figure=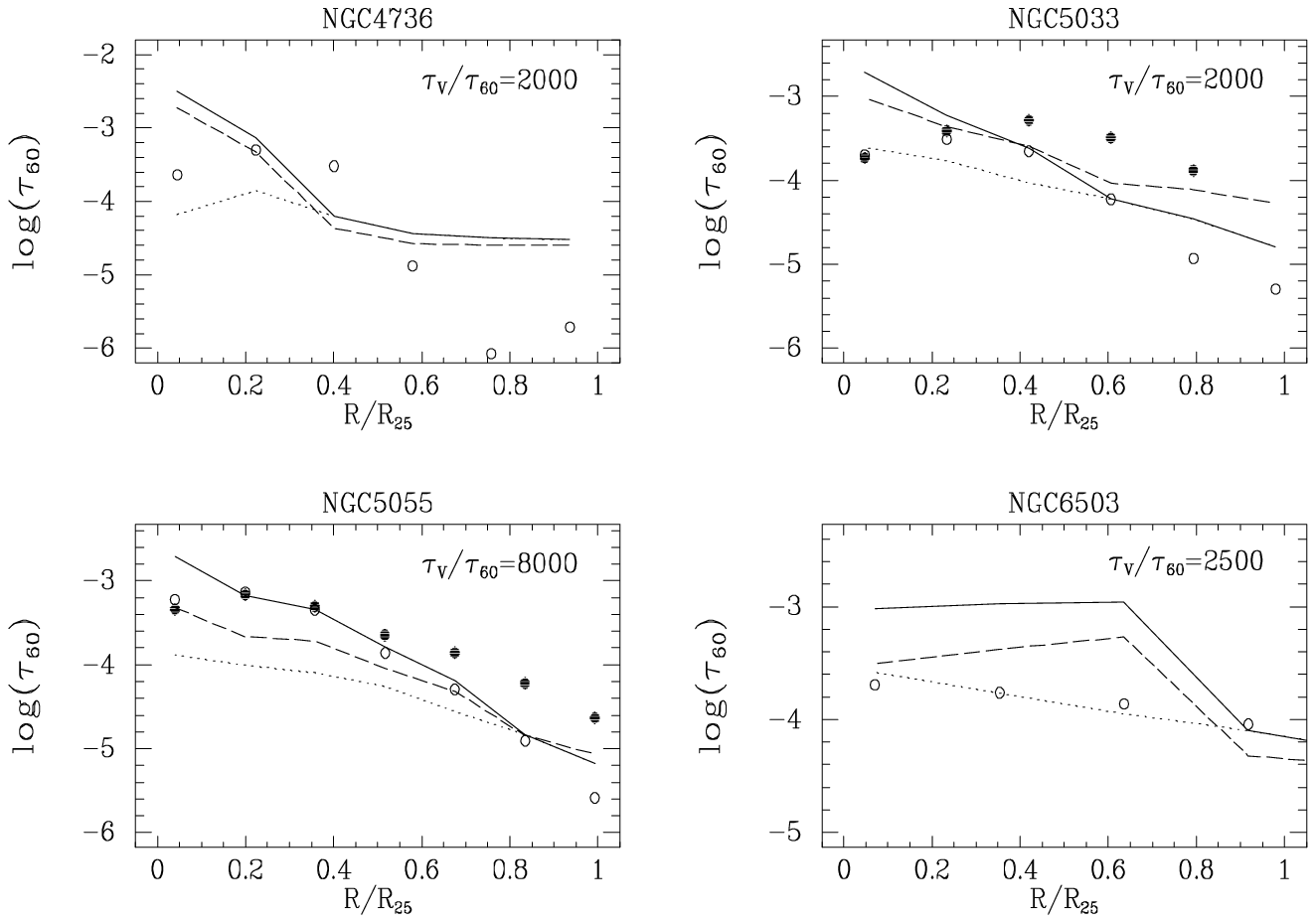,height=11.0cm}}
\vspace*{-1.0cm}
\centerline{\psfig{figure=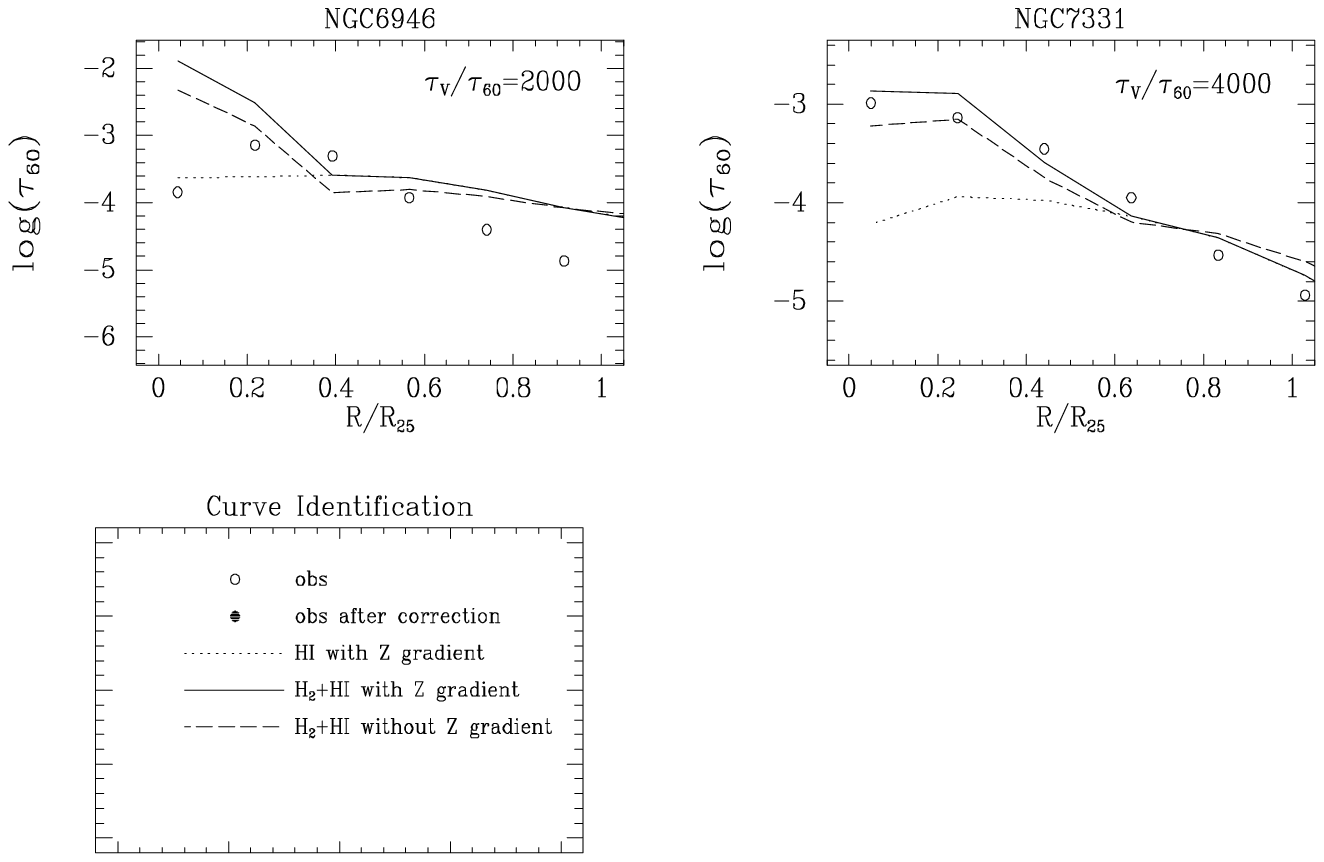,height=11.0cm}}
\vspace*{-0.5cm}
\centerline{Fig.~5.--- Continued.}
\end{figure}

\begin{figure}[htb]
\vspace*{-1.0cm}
\centerline{\psfig{figure=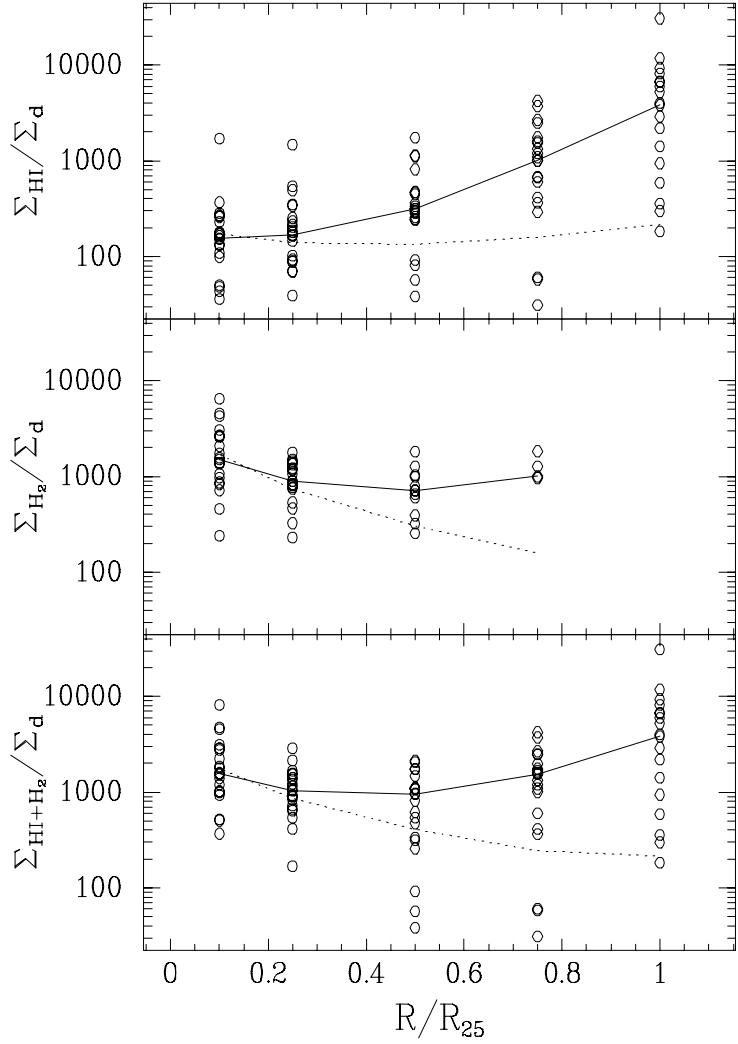,height=15.0cm}}
\vspace*{-0.0cm}
\caption{
Observed values of the gas-to-dust mass surface density ratios in the program 
galaxies are plotted at distances 10, 25, 50, 75 and 100\% of the optical 
disk radius. At each radius there is one open circle for each galaxy. 
The solid line passes through the median of the distribution. Dust
masses are corrected for the contaminating effects of VSGs and PAHs and
the resulting radial profiles are denoted by the dotted lines. 
Individual atomic and molecular gas masses in the first two panels add 
up to give the total gas-to-dust ratio of panel 3. 
}
\end{figure}

\begin{figure}[htb]
\vspace*{-2.0cm}
\centerline{\psfig{figure=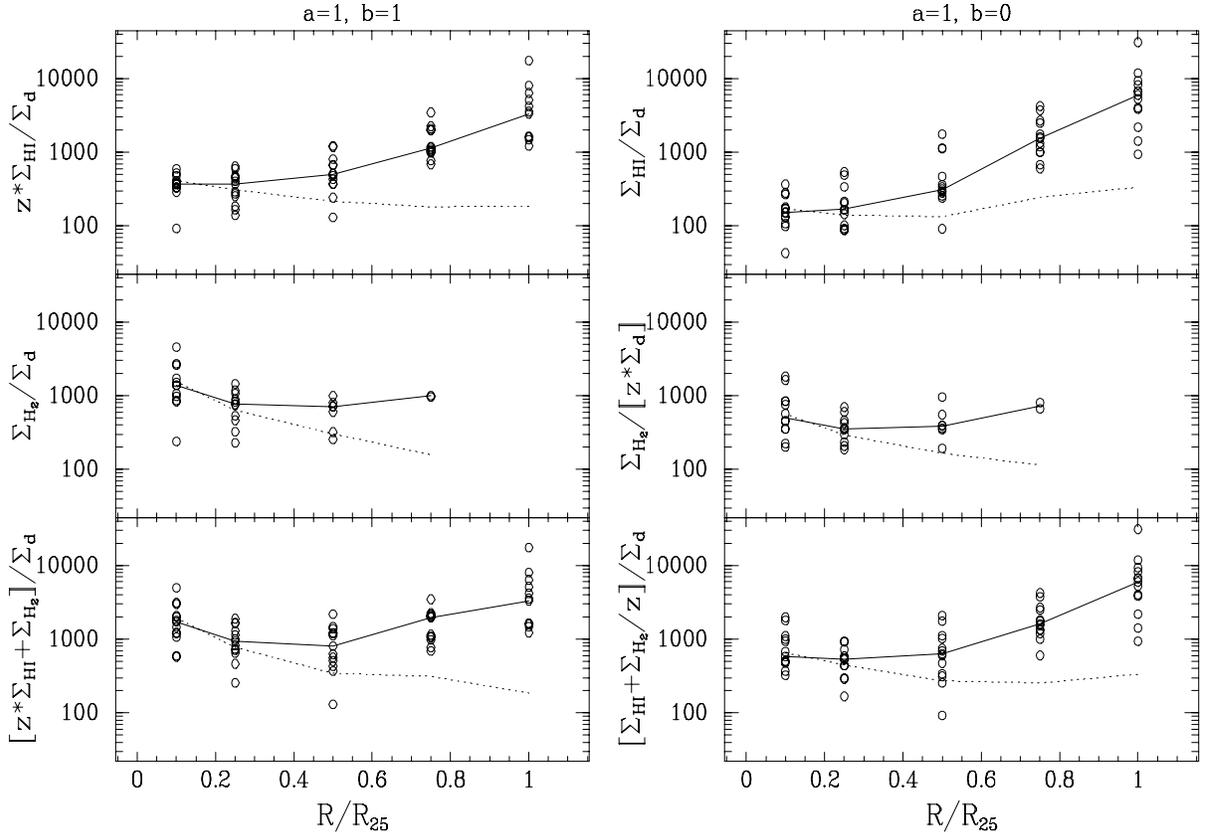,height=12.0cm}}
\vspace*{-0.0cm}
\caption{
The role of metallicity ($z$) in producing the radial gradients in
the gas-to-dust ratio is investigated in this figure. 
The CO intensity to \htwo\ mass conversion factor is assumed to depend
inversely on $z$($a=1$) in all the plots. The dust fraction is assumed to be
proportional to $z$ ($b=1$) in the left three panels. Other details are
similar to that in Fig.~6. The corrected profile is nearly
horizontal in the last panel, implying that the dust fraction is independent
of $z$, but the CO intensity to \htwo\ mass conversion factor indeed depends
inversely on $z$.
}
\end{figure}

\begin{figure}[htb]
\vspace*{-2.0cm}
\centerline{\psfig{figure=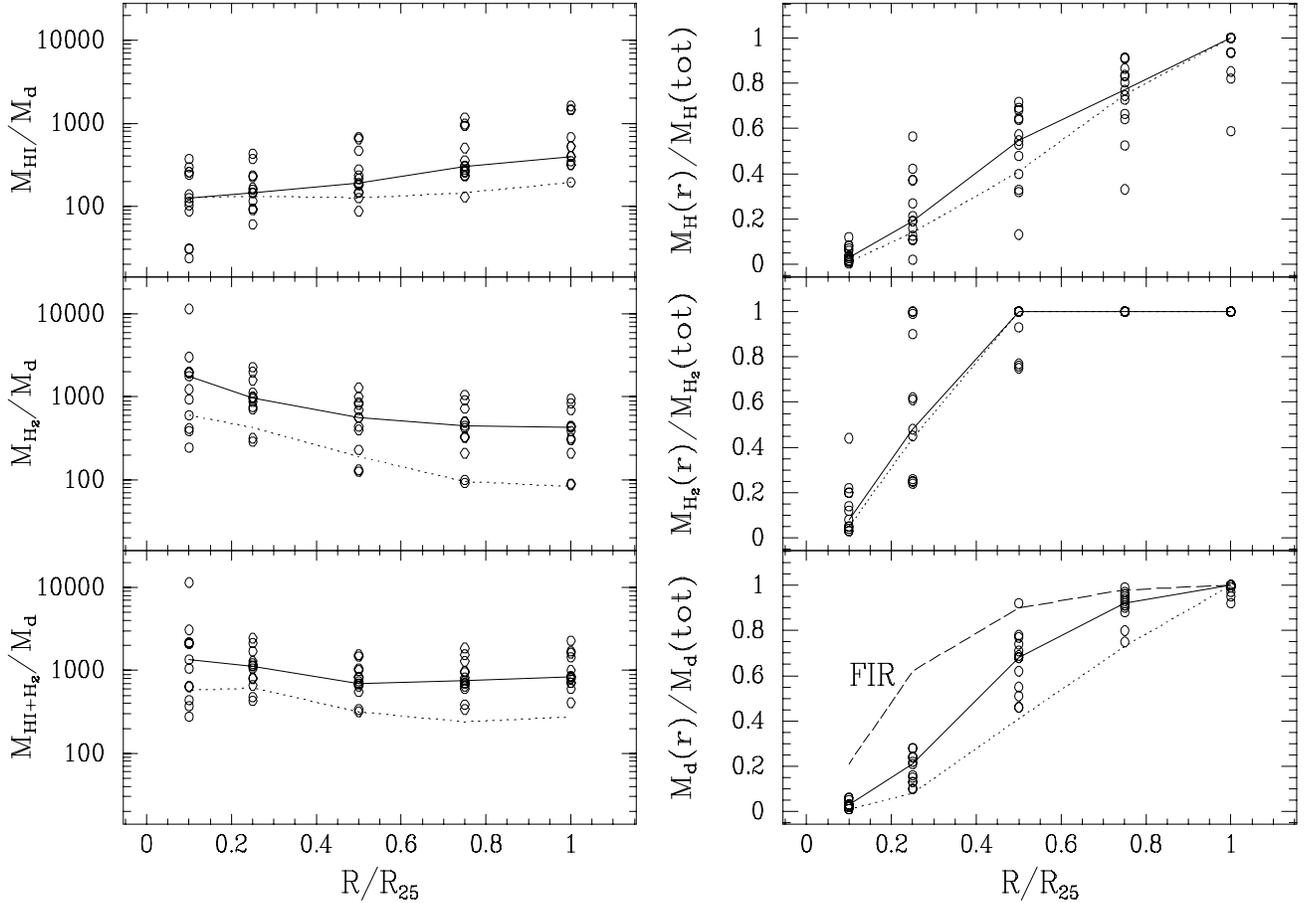,height=13.0cm}}
\vspace*{-0.0cm}
\caption{
The cumulative gas-to-dust mass ratios are plotted as a function of radial
distance from the center in the left 3 panels. The solid line and open
circles have the same meaning as in Fig.~6. The dotted line here takes into
account the correction to the molecular hydrogen mass in addition to the dust
mass. Note that the corrected global gas-to-dust ratio is 275, which is within 
a factor of 2 of the local galactic value, in spite of the bright central
regions having values close to 600. This trend of corrected global
values being controlled by the gas and dust in the outer disk can be
understood from the plots in the right panels. The cumulative fractions
of total gas, molecular gas and dust, to the global masses
in the respective components are plotted in the three panels. The
corrected values are denoted by the dotted line where as the circles and
solid lines denote the values for individual galaxies and their medians before 
corrections. The dashed line in the bottom most panel denotes the
cumulative FIR luminosity, 90\% of which originates within half the
disk radius. Note that the correction increases the mass fraction
outside half the disk radius from 30\% to 60\% for the dust and
45\% to 60\% for the gas. 
}
\end{figure}

\end{document}